\documentclass[useAMS,usenatbib]{mn2e}

\usepackage{times}    % fonts used by MNRAS
\usepackage{graphicx}
\usepackage{amsmath}
\usepackage{txfonts}
\usepackage{xspace}
\usepackage{enumerate}
\newcommand{\hip}{{\em Hipparcos}\xspace}
\newcommand{\hipcat}{{\em Hipparcos}\xspace periodic star catalogue\xspace}
\newcommand{\aavsocat}{{\em AAVSO}\xspace catalogue\xspace}

\newcommand{\ea}{{\em EA}\xspace}
\newcommand{\eb}{{\em EB}\xspace}
\newcommand{\ew}{{\em EW}\xspace}
\newcommand{\elli}{{\em ELL}\xspace}
\newcommand{\lpv}{{\em LPV}\xspace}
\newcommand{\rv}{{\em RV}\xspace}
\newcommand{\cw}{{\em CW}\xspace}
\newcommand{\cwa}{{\em CWA}\xspace}
\newcommand{\cwb}{{\em CWB}\xspace}
\newcommand{\dcep}{{\em DCEP}\xspace}
\newcommand{\dceps}{{\em DCEPS}\xspace}
\newcommand{\cepb}{{\em CEP(B)}\xspace}
\newcommand{\rrab}{{\em RRAB}\xspace}
\newcommand{\rrc}{{\em RRC}\xspace}
\newcommand{\gdor}{{\em GDOR}\xspace}
\newcommand{\dsct}{{\em DSCT}\xspace}
\newcommand{\dsctc}{{\em DSCTC}\xspace}
\newcommand{\sxphe}{{\em SXPHE}\xspace}
\newcommand{\bcep}{{\em BCEP}\xspace}
\newcommand{\spb}{{\em SPB}\xspace}
\newcommand{\be}{{\em BE}\xspace}
\newcommand{\gcas}{{\em GCAS}\xspace}
\newcommand{\acyg}{{\em ACYG}\xspace}
\newcommand{\acv}{{\em ACV}\xspace}
\newcommand{\sxari}{{\em SXARI}\xspace}
\newcommand{\by}{{\em BY}\xspace}
\newcommand{\rs}{{\em RS}\xspace}

\newcommand{\ntree}{$n_{tree}$\xspace}
\newcommand{\mtry}{$m_{try}$\xspace}

\title[Random forest classification of \hip periodic variable
stars]{Random forest automated supervised classification of
  \hip periodic variable stars}

\author[P. Dubath et al.]{P. Dubath,$^{1,2}$\thanks{E-mail: Pierre.Dubath@unige.ch} 
L. Rimoldini,$^{1,2}$ 
M. S\"uveges,$^{1,2}$
J. Blomme,$^{3}$
M. L\'opez,$^{4}$
 L. M. Sarro,$^{5}$
\newauthor
J. De Ridder,$^{3}$
 J. Cuypers,$^{6}$
L. Guy,$^{1,2}$
I. Lecoeur,$^{1,2}$
K. Nienartowicz,$^{1,2}$ 
A. Jan,$^{1,2}$
\newauthor
M. Beck,$^{1,2}$
 N. Mowlavi,$^{1,2}$
P. De Cat,$^{6}$
T. Lebzelter $^{7}$
and L. Eyer$^{1,2}$\\
$^1$Observatoire astronomique de l'Universit\'e de Gen\`eve, ch. des Maillettes 51, 1290 Versoix, Switzerland\\
$^2$ISDC Data Center For Astrophysics, ch. d'Ecogia 16, 1290 Versoix, Switzerland\\
$^3$Instituut voor Sterrenkunde, K.U.Leuven, Celestijnenlaan 200D, 3001 Leuven, Belgium\\
$^4$Centro de Astrobiolog\'ia (INTA-CSIC), Departamento de
Astrof\'isica, PO Box 78, E-28691, Villanueva de la Ca\~nada, Spain\\
$^5$ Dpt.\ de Inteligencia Artificial , UNED, Juan del Rosal, 16, 28040 Madrid, Spain \\
$^6$ Royal Observatory of Belgium, Ringlaan 3, 1180 Brussels, Belgium \\
$^7$ University of Vienna, Department of Astronomy, T\"urkenschanzstrasse 17, A1180 Vienna, Austria }

\begin{document}

\date{Accepted 2011 February 21. Received 2011 February 18; in
  original form 2011 January 12}

\pagerange{\pageref{firstpage}--\pageref{lastpage}} \pubyear{2011}
%\linenumbers
%\modulolinenumbers[5]

\maketitle

\label{firstpage}

\begin{abstract}

  We present an evaluation of the performance of an automated
  classification of the \hip periodic variable stars into 26
  types. The sub-sample with the most reliable variability types
  available in the literature is used to train supervised algorithms
  to characterize the type dependencies on a number of attributes. The
  most useful attributes evaluated with the random forest methodology
  include, in decreasing order of importance, the period, the
  amplitude, the V-I colour index, the absolute magnitude, the
  residual around the folded light-curve model, the magnitude
  distribution skewness and the amplitude of the second harmonic of
  the Fourier series model relative to that of the fundamental
  frequency. Random forests and a multi-stage scheme involving
  Bayesian network and Gaussian mixture methods lead to statistically
  equivalent results. In standard 10-fold cross-validation
  experiments, the rate of correct classification is between 90 and
  100\%, depending on the variability type. The main
  mis-classification cases, up to a rate of about 10\%, arise due to
  confusion between \spb and \acv blue variables and between eclipsing
  binaries, ellipsoidal variables and other variability types. Our
  training set and the predicted types for the other \hip periodic
  stars are available online.

\end{abstract}

\begin{keywords}
methods: data analysis -- methods: statistical  -- techniques:
photometric -- catalogues -- stars: variables: general.
\end{keywords}

\section{Introduction}

The development of efficient automated classification schemes is
becoming of prime importance in astronomy. Large surveys are
monitoring millions, and soon billions, of targets. The resulting time
series cannot possibly be scrutinized by eye. The identification and
study of variable stars require the use of powerful statistical and
data mining tools. Automated supervised classification methods provide
object type predictions based on the values of a set of attributes
characterizing the objects. In the first stage, a collection of
prototype objects of known type and attribute values, referred to as
the training set, is used to build a model of the dependencies of the
types on the attribute values. In the second stage, this model is used
to predict the types of other objects of unknown types but with
available attribute values. As the naming schemes differ in different
publications, we make the following definitions to be used throughout
this paper: {\em objects} (i.e., {\em stars}) are classified into {\em
  types} making use of a number of {\em attributes}.

Tree-based classification methods are simple to use and popular in
applications (see e.g. \citet{Hastie+09}). They can deal with complex
structures in the attribute space and have low systematic
classification errors if the trees are sufficiently deep. However
trees are noisy and the resulting type estimator has large
variance. With the random forest method \citep{Breiman2001} the
variance is drastically reduced by averaging the results of many trees
built from randomly selected subsamples of the training set
(bootstrapping). In addition, this method uses the best one of a
number of randomly selected attributes at each branching, which has
the effect of reducing the correlation between different trees and
hence improving the averaging.  The random method has been used in
many scientific domains such as bioinformatics, biology, pharmacy and
Earth sciences, confirming that it performs excellently on a broad
range of classification problems. By construction it is relatively
robust against overfitting, it is only weakly sensitive to choices of
tuning parameters, it can handle a large number of attributes, it
provides an unbiased estimate of the generalization error, and the
importance of each attribute can be estimated.

A number of variable star classification studies exploit recent or
ongoing surveys, such as (1) ASAS
\citep{Pojmanski02,Pojmanski03,EyerBlake02,EyerBlake05}, (2) OGLE
\citep{Sarro+09}, (3) MACHO \citep{Belokurov+02,Belokurov+04}. (4)
CoRoT \citep{Debosscher+09}, (5) Kepler \citep{Blomme+2010}.  A number
of ambitious survey projects are also in an advanced stage of
preparation, in particular, (1)
Pan-STARRS\footnote{http://pan-starrs.ifa.hawaii.edu/public}, (2)
LSST\footnote{http://www.lsst.org/lsst} and (3)
Gaia\footnote{http://www.rssd.esa.int/Gaia}. Although these projects
have different primary goals, the expected time series of measurements
will provide data of unprecedented quality to study variable
stars. These data are critical to both better characterize populations
of variable stars in different environments and investigate in more
depth typical or peculiar individual cases.  Progress in this field
not only impacts our understanding of variable star physics, but also
leads to contributions to a wide range of astronomical topics, from
stellar evolution and population-synthesis modelling to distance scale
related issues.

The Hipparcos mission stands out as a rather original and ambitious
astrometry space programme. Because of the whole sky repeated
scanning, it provides accurate data for all the brightest stars in our
close neighborhood. The \hipcat includes most of the best studied
stars and hence provides a unique set that can be used as a
``control sample''. Results obtained for these stars can be
validated using the wealth of available published information. This is
particularly useful for evaluating variable star classification
methods before applying them to other large surveys. The variability
types resulting from the classification can be compared with types
available from the literature. The \hip sample also includes almost
all types of variable stars present in the solar vicinity. It is
certainly a solid basis for building a training sample for supervised
classification methods.

The comprehensive study of \hip variable stars presented in volume 11
of the
\hipcat\footnote{http://www.rssd.esa.int/index.php?project=HIPPARCOS\&page=Overview}
\citep{Eyer98phd} does not comprise a systematic automated
classification. The variability types provided in this catalogue are
extracted from the literature with two exceptions. First, the
eclipsing binaries and the RV Tauri were identified and characterized
on the basis of visual inspections of the folded light curves. Second,
a systematic classification of variables with B spectral type was
achieved by \cite{Waelkens+98} using a multivariate discriminant
analysis. Later, \cite{Aerts+98} used a similar technique to isolate
variables with A2 to F8 spectral types. An attempt to obtain a
systematic classification is also presented by \cite{WillemsenEyer07}.

% Some automated classifications
% or extractions have been achieved for particular types of
% variables. For example,  \cite{Waelkens+98} classify B variable stars
% with a Multivariate Discriminant Analysis,  while \cite{Aerts+98} use a
% similar technique to isolate variables with A2 to F8 spectral
% types. An attempt to obtain a systematic classification is also
% presented by \cite{WillemsenEyer07}.

The subject of this paper is the development of the first systematic,
fully automated classification of the complete sample of \hip periodic
variable stars. The performance of the random forest method is
evaluated and the results are compared with those obtained from the
multistage classifier developed recently \citep{Blomme+2011}. In a
companion paper \citep{Rimoldini+2011}, this work is extended to
include non-periodic variables, both to study the classification of
non-periodic stars and to evaluate the confusion between periodic and
non-periodic types. The main goal is to fully validate our approach on
a controlled sample of \hip stars, before applying it in a more
automated fashion on other surveys in future studies. An important
outcome of our work is a homogeneous supervised classification
training set, which can be adapted to other missions, in particular,
the upcoming Gaia mission. Predicted types are also provided for
almost all \hip periodic variables, including some that have no types
or only uncertain ones in the literature.

The procedure followed to build the training set from a sub-sample of
the best-known \hip stars is described in Sect.~\ref{SECT_TS}. The
subject of Sect.~\ref{SECT_ATTR} is the determination of the attribute
values, including the period search. Section~\ref{SECT_RF} describes
the applied random forest methodology and shows the corresponding
results. Section~\ref{SECT_PERR} presents an investigation of the
influence of the period value errors on the classification
process. Results from random forest results and a multi-stage
classifier are compared in Sect.~\ref{SECT_MULT}, while our best final
predicted types are listed in Sect.~\ref{SECT_TYPES}. Finally, our
conclusions are given in the last section (\ref{SECT_CONC}).

\section[]{Training set composition \label{SECT_TS}}

Data from a set of objects of known types are needed to train the
supervised classification algorithms. The quality of the
classification directly rests on the reliability of the types of the
stars included in this set, called the training set. For a given type,
the selected objects should only include true representatives of the
group with typical properties. In our case, we select a sub-set of our
\hip stars with most reliable types available from the literature,
taking advantage of the fact that many of them are relatively bright,
well studied objects. Ideally, the relative frequencies of the
different types in the training set should be representative of those
in the population to-be-classified.

A search for periods in the Hipparcos data alone is inconclusive for
171 of the 2712 stars included in the \hipcat due to the incomplete
phase coverage of the light curves.  The period values published in
the \hipcat come from the literature for these stars. Most of them are
eclipsing binaries (152 EAs) with too few Hipparcos measurements
during the eclipses. These stars are excluded from the training set as
the scope of this paper is restricted to periodic stars with a
light curve from which it is possible, at least in principle, to infer
a period.

The variability types provided in the \hipcat were mainly extracted
from the literature (to the notable exception of eclipsing binaries,
see Sect.~\ref{SECT_PER_ECL}) available at the time of publication
(1997).  These types are revised for our study using more recent
information. The main reference is the International Variable Star
Index \citep{aavso_cat} catalogue from the American Association of
Variable Star Observers (\aavsocat hereafter). This index includes
information from the General Catalogue of Variable Stars (GCVS) and
the New Catalogue of Suspected Variables (NSV) and it is kept
up-to-date with the literature with 2 releases per month. The release
adopted herein is that from June 13th, 2010. Information from private
communications is preferentially used for some specific variability
types (see below) as it is believed to be more reliable.
% (see Table~\ref{TAB_TS}).

The type-assignment process for \hip variables is as follows (see
Table~\ref{TAB_TS} for type acronym definitions):

\begin{enumerate}[1.]
\item For eclipsing binaries and ellipsoidal variables, the \hipcat is
  taken as the reference as (1) the classification done at the time
  included reliable visual checking of the Hipparcos light curves and
  (2) these light curves are known to have a good enough eclipse
  coverage to allow a successful period determination (see above and
  Sect.~\ref{SECT_PER_ECL}).

\item Lists of \hip stars of the types \gdor, \spb and \bcep are
  provided by P. De Cat and of \lpv by T. Lebzelter, both maintain
  up-to-date compilations of literature information for these types.

\item The type determination for \acv and \sxari stars that have a
  measured magnetic field are considered as particularly reliable. As
  a consequence, for these types, only the \hip stars included in
  a list of magnetic stars provided by I.I. Romanyuk (private
  communication) are retained.

\item Only the subset of \hip ~\rs and \by stars listed in the third
  edition of the ``catalogue of chromospherically active binary
  stars'' \citep{CAB2008} are included.

\item All stars from the \aavsocat with a type matching any of the above
  mentioned types are excluded. For example, a star identified as Mira, SR, LB or
  SARV in \aavsocat that is not in the Lebzelter list of \lpv is
  discarded from the training set. The \aavsocat is then used to
  assign a type to the remaining stars from the \hipcat. This
  procedure leads to a subset of 1963 stars.
 %(out of the total of the  2712 - 171  = 2541 variables).

\item A visual inspection of the folded light curves of all 1963 stars
  leads to the elimination of 64 stars due to either poor sampling or
  excessive noise.

\item Types with less than three representatives are discarded. This
  concerns 7 stars of six different types: FKCOM, HADS, INSA, INSB,
  nra, CW-FU in \citet{aavso_cat}.

\item Finally, 92 stars with uncertain type (denoted with a colon in
  the original sources) are also excluded.

\end{enumerate}

\begin{table}
 \begin{center}
  \caption{\label{TAB_TS}  Training set composition }
  \begin{tabular}{llrl}
  \hline
\multicolumn{2}{l}{Type}     &     \multicolumn{1}{c}{Num}    & Main reference   \\
 \hline
Eclipsing Binary                      &\ea        &   228    & Hipparcos \\
                                             &\eb        &   255    & Hipparcos \\
                                             &\ew        &   107    & Hipparcos \\
Ellipsoidal                               &\elli        &   27    & Hipparcos \\
Long Period Variable              &\lpv        &   285    & Lebzelter (p. c.) \\
RV Tauri                                & \rv          &     5     &  AAVSO\\
W Virginis                              &\cwa       &     9    &  AAVSO  \\
                                              &\cwb       &     6    & AAVSO \\
Delta Cepheid                      &\dcep         &    189   & AAVSO \\
~~~(first overtone)                     &\dceps  &     31    & AAVSO \\
~~~(multi mode)                      &\cepb      &     11   & AAVSO \\
RR Lyrae                                   &\rrab       &      72   & AAVSO \\
                                                &\rrc         &       20  & AAVSO \\
Gamma Doradus                      &\gdor       &       27  & De Cat (p. c.) \\
Delta Scuti                              &\dsct       &       43 & AAVSO \\
~~~(low amplitude)                  &\dsctc      &     81    & AAVSO \\
SX Phoenicis                           &\sxphe      &    4     & AAVSO \\
Beta Cephei                            &\bcep       &     30   & De Cat (p. c.) \\
Slowly Pulsating B star             &\spb        &     81   & De Cat (p. c.) \\
B emmission line star              &\be         &       9   & AAVSO \\
Gamma Cassiopeiae                &\gcas       &      4     & AAVSO \\
Alpha Cygni                             &\acyg       &     18  & AAVSO \\
Alpha-2 Canum Venaticorum &\acv        &   77      & Romanyuk  (p. c.) \\
SX Arietis                                 &\sxari      &     7     & Romanyuk  (p. c.) \\
BY Draconis                              &\by         &      5     &  \cite{CAB2008} \\
RS Canum Venaticorum          &\rs          &    30     & \cite{CAB2008} \\
\hline 
\multicolumn{2}{r}{\bf Total:}  & 1661 & \\
\hline 
\end{tabular} \\
\end{center}
 AAVSO : \cite{aavso_cat} \\
\end{table}

The above type-assignment process leads to a training set with 1800
stars of 26 different types. A further 32 stars are excluded due to
diverse difficulties in the light-curve processing (see
Sect.~\ref{SECT_ATTR}) and another 107 are discarded because of
missing colour indices in the \hipcat.  This leaves a training set of
1661 stars.  Table~\ref{TAB_TS} summaries the final composition of the
training set as a function of the variability type.

It is important to note that combined types, such as an intrinsic
variable included in an eclipsing binary, are excluded from our
training set as a result of the above type assignment process.

\section[]{Classification attributes \label{SECT_ATTR}}

The classification experiments presented in this paper rely on a
number of attributes.  To achieve the most accurate classification,
attributes should be chosen so as to characterize the stars as
thoroughly as possible.  Some attributes reflect stellar global
properties, such as the mean colour or the absolute brightness whereas
others describe features of the light curve. The shape of the folded
light curve is one of the key indicators of variability type.

The strategy used in this paper is to compute a large number of
attributes and to use some algorithms to estimate their merits. In
this section, we describe the principle of the attribute
derivation. The attribute ranking and selection is described in
Sect.~\ref{SECT_RF}, which is devoted to classification. The exact attribute
definition is also deferred to avoid describing some that may finally
not be retained.

\subsection{Statistical parameters}

A number of statistical parameters are derived from the distribution
of photometric measurements. The list includes the distribution
moments (mean, standard deviation, skewness and kurtosis), the range
and percentiles. Weighted and un-weighted formulations are used as
well as robust estimators.

\subsection{Period search \label{SECT_PER}}

The period values provided in the \hipcat are particularly reliable as
the corresponding folded light curves were all visually checked prior
to publication. Using directly these values in our analysis would lead
to optimum results. However, current and upcoming surveys can include
millions of variables which cannot all be visually checked. One of the
main goals of this paper is to investigate what can be achieved
through an {\em automated} classification process, including the
period search\footnote{The problem of spurious and aliased periods (or
  frequencies) typically showing up at fractions and multiples of one
  day and one year in ground based surveys does not affect the period
  search in Hipparcos data.}. An investigation of the increase of the
classification errors resulting from the use of an incorrect period is
presented in Sect.~\ref{SECT_PERR}.

A number of well-known period search methods such as
\citet{Deeming1975}, Lomb-Scargle \citep{Lomb1976, Scargle1982},
harmonic least squares analysis of generalized Lomb-Scargle methods
\citep{Zechmeister+2009}, String Length methods \citep{Lafler+1965,
  Burke+1970, Renson1978, Dworetsky1983} and Jurkevich-Stellingwerf
\citep{ Jurkevich1971, Stellingwerf1978} are employed to search for
periodicity in the light curves. The resulting periods are compared
with the \hipcat values to derive the fraction of correct
results. Extensive testing shows that a single method can lead to a
recovery fraction of around 80\%, while an ideal combination of all
methods could potentially raise that value to close to 100\% (Cuypers
2011, in preparation). Unfortunately, no automated strategy is found
to predict which method leads to the correct period for a specific
light curve. The best overall recovery fraction is obtained with a
combination of the classic Lomb-Scargle method \citep{Lomb1976,
  Scargle1982} and of a generalised Lomb-Scargle variant
\citep{Zechmeister+2009}. The former is used for all stars with a
large magnitude distribution skewness while the latter is used for all
remaining stars.

The period search results obtained for eclipsing binaries and for the
other types of variables are different. They are presented separately
in the next two sub-sections. 

\subsubsection{Non-eclipsing variable periods \label{SECT_PER_OTHER}}

\begin{figure}
 \includegraphics[width=\columnwidth]{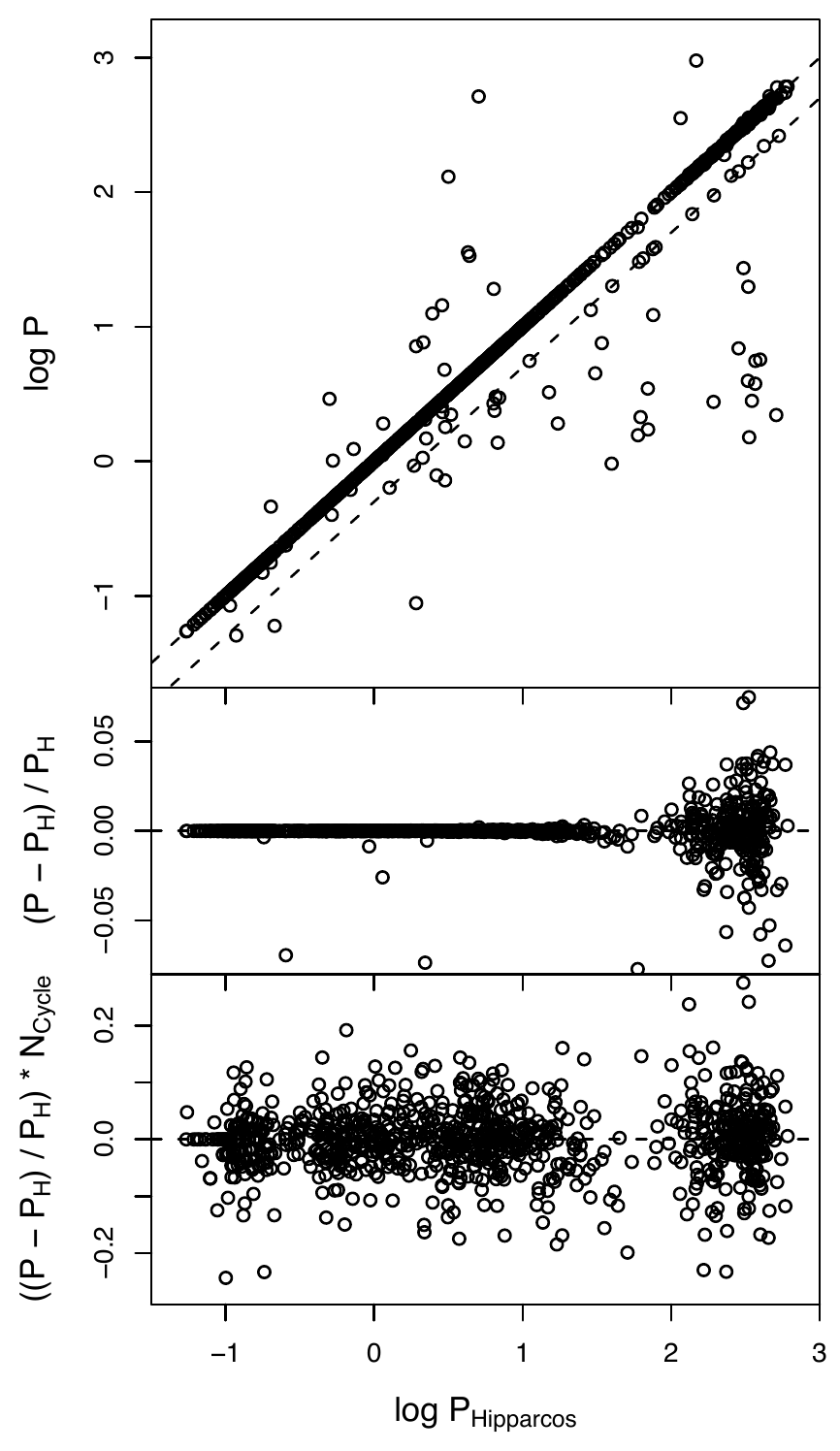}
 \caption{Periods extracted using the Lomb-Scargle method for
   non-eclipsing variables as a function of the \hipcat periods $P_H$
   expressed in days on a decadic log scale. The upper plot shows the
   Lomb-Scargle period (P), the middle one the relative difference,
   and the lower one the relative difference multiplied by
   N$_{Cycle}$. N$_{Cycle}$ is the light curve span divided by the
   period value, i.e., the number of cycles between the beginning and
   the end of the light curve. The lower diagram shows the cumulative
   shift, expressed in units of phase, after N$_{Cycle}$ resulting
   from the inaccuracy of the period value. The dashed diagonal lines
   in the upper plot display the relationships P = P$_{\rm H}$, and P
   = 0.5 P$_{\rm H}$. The middle and lower plots have much enlarged
   y-scales so that almost all outliers visibly scattered in the upper
   plot fall outside of the displayed ranges.}
\label{FIG_PER_OTHER}
\end{figure}

Figure~\ref{FIG_PER_OTHER} displays the results of the period search
for all variables excluding the eclipsing binaries and the
ellipsoidals using the classic or generalised Lomb-Scargle methods
(see above Sect.~\ref{SECT_PER}) depending on the skewness value. Out
of the 1044 non-eclipsing variables, a good period estimate is derived
for 951 stars, i.e., a correct period recovery rate of 91\%.  The
period is considered as good if the difference between the extracted
period and the \hip catalogue value does not lead to a cumulative
shift in phase of more than 20\% over the full time span of the light
curve.

\subsubsection[]{Eclipsing variable periods \label{SECT_PER_ECL}}

The most common type of photometric variability is due simply to
binary stars eclipsing each other. This represents a real challenge
for classification because almost all kinds of stars can form a binary
system. For example, the colour index, usually a powerful indicator of
stellar type, is no longer a useful discriminant as it is a mean from
two stars that can have almost any kind of stellar colours. In
addition, one of the stars, or even both, can exhibit other types of
variability leading to a wide range of combined behaviours. Close
interaction can also trigger other types of variability such as the RS
Canum Venaticorum phenomenon.

\begin{figure}
 \includegraphics[width=\columnwidth]{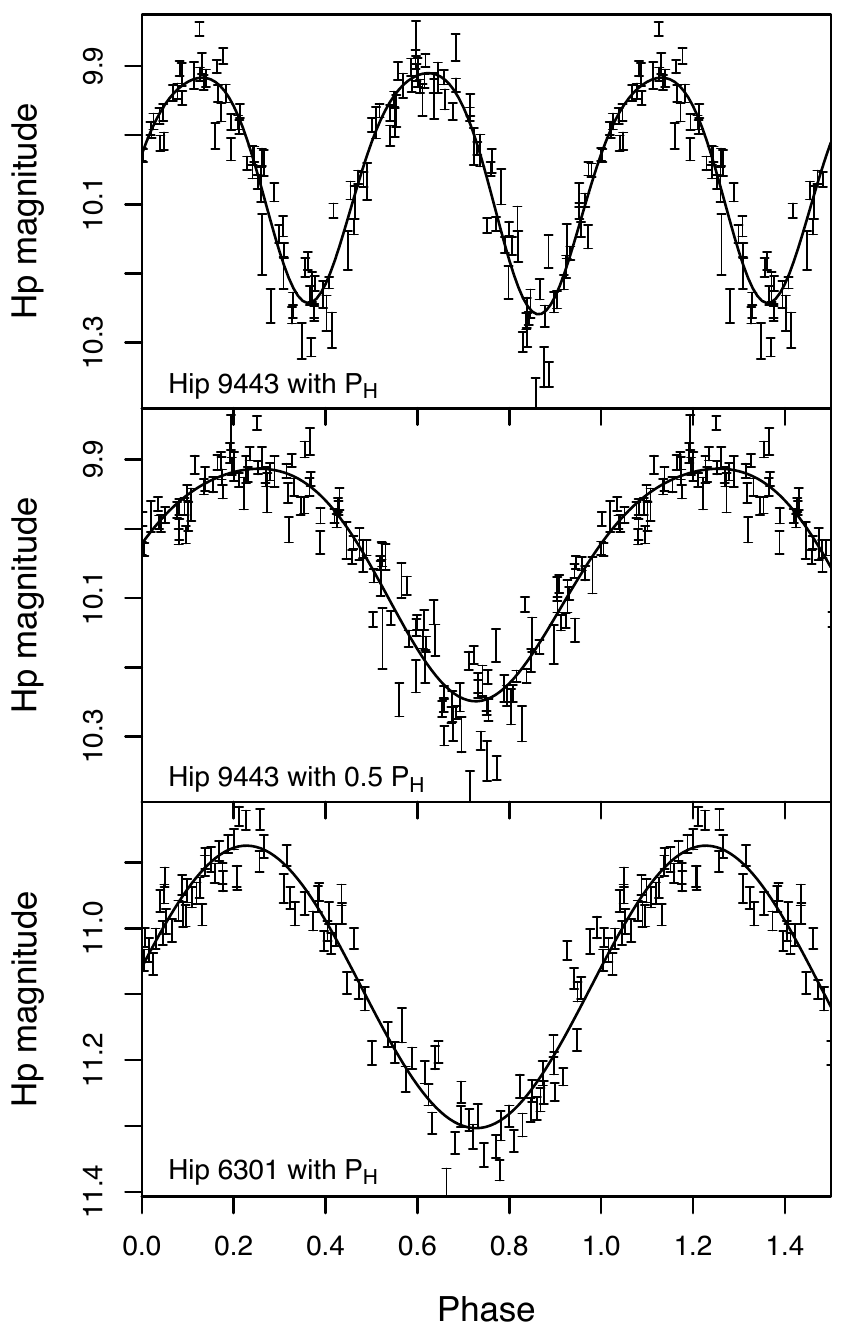}
 \caption{The top and bottom panels show that the folded light curves of
   the \eb Hip 9443 and of the \rrc Hip 6301 can be easily
   differentiated. However, an automatic period search is most likely
   to result in half of the correct period for Hip 9443 leading to the
   folded light curve displayed in the middle panel. This curve is
   similar to that of the bottom panel illustrating the difficulties
   of distinguishing these two stars in an automated process.}
\label{FIG_LC_ECL}
\end{figure}

There is an important complication in deriving the periods of
eclipsing binaries. The folded light curves (or the pulse profile) of
non-eclipsing periodic variable stars exhibit usually a single
excursion per cycle going through a unique minimum and maximum.  As
two eclipses are often observed over one binary system revolution,
the resulting light curves exhibit two minima over one cycle. In
significant number of cases, the two minima have almost the same depth
and width. The light curves exhibit two almost identical excursions
and consequently, the period search usually returns half of the
true period. Figure~\ref{FIG_LC_ECL} shows examples of actual
Hipparcos light curves to illustrate the difficulty in extracting the
correct period for eclipsing binary systems.

In the preparation of the \hipcat, the light curves of all the 2712
periodic variables were visually inspected. Making sometimes use of
additional information from the literature, the eclipsing binaries
were identified and when necessary the period doubled. Introducing
these period values into a general classification algorithm is not
appropriate. Because of the double-excursion behaviour of their
light curves they could be very easily separated from the other
variables. But this success would be an illusion as it simply reflects
the fact that these eclipsing binaries were carefully identified and
their period confirmed through a thorough visual check in the first
place.

\begin{figure}
 \includegraphics[width=\columnwidth]{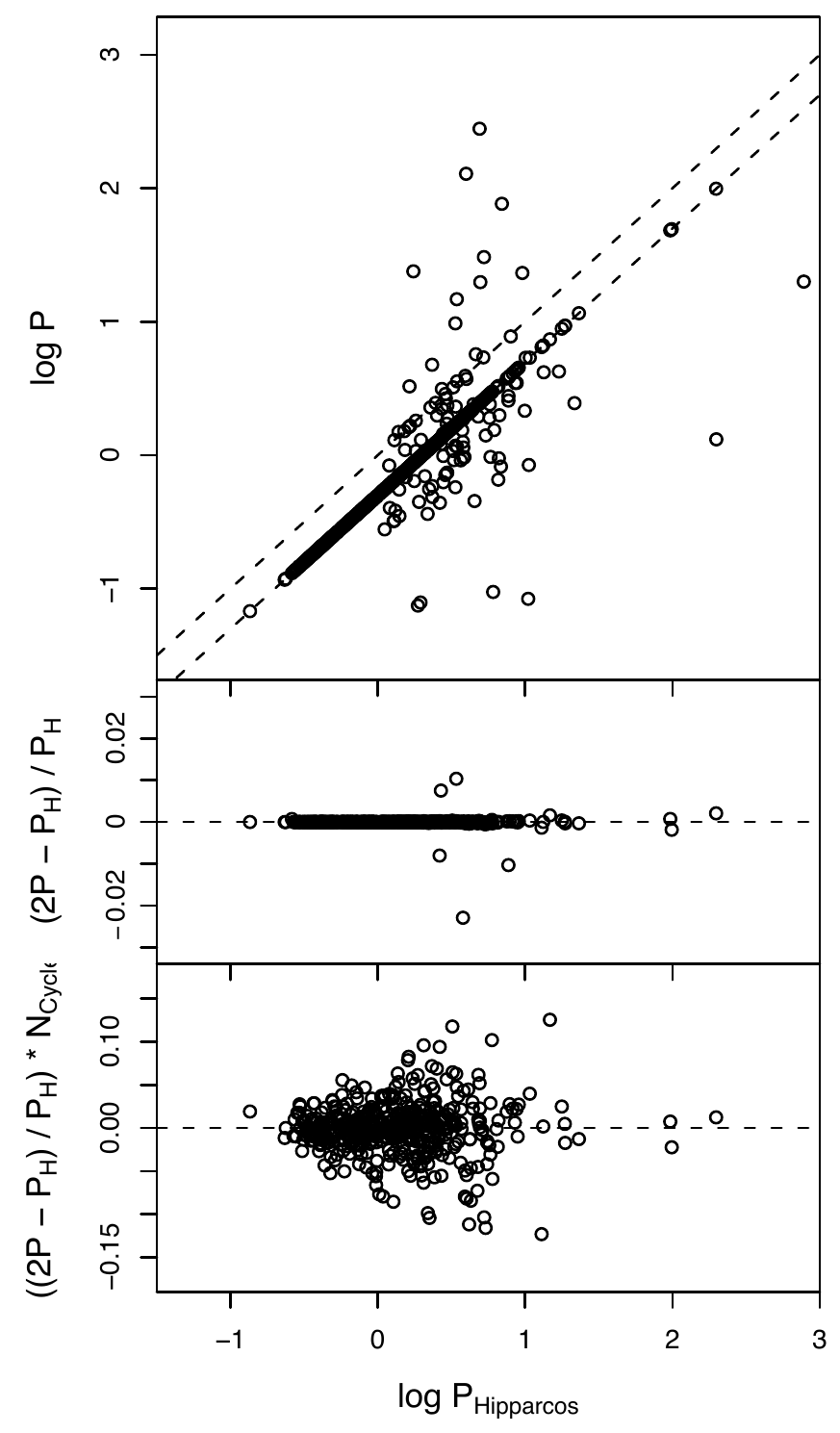}
 \caption{Periods extracted using the Lomb-Scargle method for
   eclipsing variables as a function of the \hipcat periods $P_H$
   expressed in days on a decadic log scale. The upper plot shows the
   Lomb-Scargle period (P), the middle one the relative difference,
   and the lower one the relative difference multiplied by
   N$_{Cycle}$. N$_{Cycle}$ is the light curve span divided by the period
   value, i.e., the number of cycles between the beginning and the end
   of the light curve. The lower diagram shows the cumulative shift,
   expressed in unit of phase, after N$_{Cycle}$ resulting from the
   inaccuracy of the period value. The dashed diagonal lines in the
   upper plot display the relationships P = P$_{\rm H}$, and P = 0.5
   P$_{\rm H}$. The middle and lower plots have much enlarged y-scales
   so that almost all outliers visibly scattered in the upper plot
   fall outside of the displayed ranges.  }
   \label{FIG_PER_ECL} 
\end{figure}

Figure~\ref{FIG_PER_ECL} displays the results of the period search for
the 617 eclipsing binaries (EA, EB, and EW) and ellipsoidal variables
using the classic or generalised Lomb-Scargle methods (see above
Sect.~\ref{SECT_PER}) depending on the skewness value. Out of the 617
variables, approximately half of the \hip period is obtained for 508
stars, i.e., for 82\% of the sample. For these 508 cases, the
difference between the double of the extracted period and the
Hipparcos value does not lead to a cumulative phase shift of more than
15\% over the full time span of the light curve.  A period close to
the \hip full period value is obtained only for 10 cases. Some period
search algorithms are better than Lomb-Scargle to find the correct
period. With Jurkevich-Stellingwerf, for example, the correct value
and half of it are obtained in similar numbers of cases (38\%).  The
trouble is that when the true period is unknown, it is impossible to
know in which cases, the double or the full value of the extracted
period is the correct result. For this paper, we have found that the
best strategy is to use the Lomb-Scargle period to model the light
curves and to double the period values after classification for all
objects that have been identified as eclipsing binaries and
ellipsoidal variables. In this way, almost all these stars are modeled
with half of the correct periods, as shown in the middle panel of
Fig.~\ref{FIG_LC_ECL}. The use of half the true period in the case of
eclipsing binaries does not much confuse the classifier and leads to
the best overall results.

\subsection{Light curve modelling}

The \hipcat provides a unique period for each source. Although a
number of these sources are truly multi-periodic, looking at the
folded light curves displayed in the \hipcat shows that in the vast
majority of cases the curves obtained with the single, dominant period
look good. Indications of additional significant periods, such as an
apparent superposition of two curves or a strong scatter excess with
respect to the nominal photometric uncertainties, are only evident in
few cases. In this paper, we show that a light-curve modelling carried
out with the dominant period is sufficient to achieve a reliable
classification.

The time-series model is given by:

\begin{equation}
y = a_0 + \sum_{k=1}^{N_{h}} 
      b_{k} \cos(2\pi k \nu t) + c_{k} \sin(2\pi k \nu t),
\label{tsmodel1} 
\end{equation} 

\noindent where 
% we assume that the reference epoch $t_{ref}$ has already been
% subtracted from the time values.
$N_{h}$ is the number of significant harmonics. Here, $k=1$ is the
fundamental frequency $\nu$ (or first harmonic) and $k=2$ is the first
overtone (or second harmonic). This model is linear in its
coefficients $\bbeta = \{a_0,b_{k},c_{k}\}$. An alternative notation
involves the amplitudes and phases of the oscillations and their
overtones

\begin{equation}
y = a_0 + \sum_{k=1}^{N_{h}} 
      A_{k} \sin(2\pi k \nu t + \varphi_{k}).
\label{tsmodel2} 
\end{equation} 
      
\noindent The coefficient correspondence is

\begin{eqnarray}
A_{k} & = & \sqrt{b_{k}^{2} + c_{k}^{2}} \\
\varphi_{k} & = & \arctan(b_{k},c_{k}).
\end{eqnarray}

The number of harmonics ($N_{h}$) to be used to best fit a given light
curve is unknown and must be determined through the modelling
process. A model with more parameters, i.e., a higher number of
harmonics, will always fit the data at least as well as the model with
fewer harmonics. The question is whether the model with additional
harmonics gives a {\em significantly} better fit to the data. A {\em
  forward selection} regression process is followed, starting with
only the fundamental coefficients, i.e., with $N_{h}$ = 1 and adding
one more harmonic at a time. The different models are {\em nested},
i.e., a simpler model can be obtained by zeroing one or more
coefficients of a more complex model.  Different models are compared
pairwise using an $F$-test.  For example, in order to compare model 1
and model 2, having different harmonic numbers ($N_{h2} > N_{h1}$),
the following $F$ statistic is computed

\begin{equation}
f = \frac{RSS_1 - RSS_2} {RSS_2} \; \frac{n_{obs} -P_{2}}
{P_{2} - P_{1}},
\label{f-test} 
\end{equation} 

\noindent where $RSS$ are the residual sum of squares, $n_{obs}$ is
the number of data points (i.e., of observations), $P$ is the number
of free parameters ($P = 2 N_{h} + 1$), and where subscripts 1
and 2 refer to models 1 and 2, respectively . Under the null
hypothesis model 2 does not provide a significantly better fit than
model 1 and $f$ follows an $F$-distribution, with ($P_{2} - P_{1}) /
(n_{obs} -P_{2}$) degrees of freedom. The null hypothesis is rejected
if the $f$ calculated from the data is greater than the critical value
of the $F$-distribution for some specified false-rejection
probability.  A succession of models with a number of harmonics ranging
from 1 to $N_{max}$ are computed and compared in turn. The $N_{h}$-th
harmonic is only retained if the corresponding model is significantly
better than the model with the lower harmonic number. The threshold to
keep a given model is $\alpha / N_{max}$ (Bonferroni correction) where
$\alpha$, set to 5\% in this study, is the overall type-I error rate
(i.e., accepting unduly one of the harmonics). There may be gaps in the
harmonic sequence when some lower harmonics are rejected while higher
ones are retained (for example, a model with $N_{h}$+2 may be better
than the model with $N_{h}$ while model with $N_{h}$+1 was rejected).

The sampling of the \hip time series is irregular with occasional
large time gaps. In some cases, models with high harmonic number which
fit the data very well exhibit large, unphysical excursions within the
gaps. In order to prevent such cases, a stop criterium based on the
width of the maximum phase gap of the time series is
introduced. The addition of new harmonics is stopped when

\begin{equation}
N_h \geq \frac{C}{2 P_{gap}},
\label{gap-test} 
\end{equation}

\noindent i.e., $N_{max}$, the maximum number of harmonics, is taken
as the first harmonic number $N_h$ in the increasing sequence that
satisfies the above inequality.  $P_{gap}$ is the width of the maximum
phase gap (between 0 and 1), and $C$ is a constant whose appropriate
value is derived empirically from extensive testing.  A optimum of 1.4
determined through visual inspections is obtained, but changing this
value in a wide range, i.e. from 1.2 to 1.6, only lead to
modifications in a handful of the \hip light-curve models.

Extensive visual inspection of the resulting models shows that this
approach is robust and that it provides good models for almost all
cases. The only notable exception is for some of the eclipsing
binaries of type \ea and \eb, with sharp eclipses, where the number of
harmonics is not sufficiently high to fully model the eclipses. In
some cases, the number of points in the eclipse are too few to provide
sufficient weight in the fitting process.There may also be phase gaps
larger than the eclipse duration. In this second case, application of
the phase gap criterium stops the iterative process and the optimum
harmonic number cannot be reached. However, this is deemed less
harmful than the large model artifacts that can occur within the gaps
if the stop criteria is not applied.

\section{Random forest \label{SECT_RF}}

Random forest \citep{Breiman2001} is a tree-based classification
method. Extensive documentation and Fortran programs by Breiman and
Cutler are available at {\it
  http://www.stat.berkeley.edu/$\sim$breiman/RandomForests/}. Both the R
randomForest package \citep{rfRpackage} and the weka \citep{weka}
implementations are used in this work.

\subsection{Algorithm \label{SECT_ALGO}}

The random forest algorithm aggregates the results of a number
(\ntree) of classification trees. Each tree is built as follows:

\begin{enumerate}[1.]
\item A {\em bootstrap} star sample is obtained by drawing a sample
  with replacements from the training set. The bootstrap sample has
  the same size as the original set, but some stars are represented
  multiple times, while others are left out. The omitted stars, called
  {\em Out-Of-Bag (OOB)}, can be used to estimate the prediction error
  (see below).

\item The tree is grown by recursively partitioning the bootstrap
  sample into subgroups with more and more homogeneous type
  content. At each node, \mtry divisions into two groups are
  considered, each using one attribute from a randomly selected set of \mtry
  attributes. The best split is selected and the process is repeated
  for the child nodes with a new set of \mtry attributes at each node.

\item A so-called {\em maximum} tree is constructed, i.e., a tree with
  terminal nodes containing only a single type of stars (or a single
  star in extreme cases).
\end{enumerate}

Typically, large numbers (500-10,000) of trees are built. Each tree
provides a predicted type for a star. The most probable type is
simply the most frequent type in the sample of predictions of the
different trees. 

An estimate of the error rate can be obtained from the training
set. Any training set star is OOB in some fraction (about one-third)
of the trees. The most frequent type obtained from all the trees where
a star is OOB provides a predicted type for each star. Note that the
sample of trees in which a given star is OOB is different for each
star. The error rate and confusion matrix can be built by comparing
the predicted with the actual types. This is similar to a
cross-validation performance estimate, but at a much lower
computational cost.

It is known that random forest is insensitive to the precise value of
\mtry. In this paper, \mtry is taken by default as the recommended
value, i.e., the square root of the number of considered attributes,
unless specified explicitly.

\subsection{Attribute importance \label{SECT_ATTR_IMP}}

Random forest can produce an attribute importance score based on the
following idea. The classification accuracy computed by passing the
OOB sample down a specific tree is recorded. The values of a given
attribute are permuted in the OOB sample, i.e., the value for a star
is randomly taken out of the sample of all other star values. The
classification accuracy is computed again with the OOB sample with
permuted values for one attribute. If this attribute is important, the
permutation should noticeably degrade the classification
accuracy. Conversely, it should not change significantly the
predictions if this attribute is ineffective in the classification
process in the first place. The attribute importance is given by the
difference in classification error averaged over all trees and
normalized by the standard deviation (of these differences). These
importance values are extensively used in our attribute selection
scheme.

\subsection{Attribute correlation and selection \label{SECT_ATTR_CORR}}

Many of the derived star attributes are highly correlated. As
explained in Sect.~\ref{SECT_ATTR}, there may be several
alternative ways to characterise a given physical property.  For
example, there are different ways to measure the amplitude of a
light curve, and different colour indices are all, to first
order, a measure of the star effective temperature. The idea is now to
investigate which of these alternatives leads to the best results in
classification.

The attribute importance estimates provided by random forest can be
used to rank attributes. The limitation is that it is not sensitive to
correlations. Two highly correlated attributes will score equally highly
in this process. Experience shows that random forest classification
results are not much affected by the use of some almost redundant,
highly-correlated attributes, but it is interesting to investigate
what is the minimum set of attributes to be used for an optimum
classification of our stars.

The recursive procedure to build a list of the most important,
not too correlated attributes is as follows:

\begin{enumerate}[1.]
\item A ranked list of attributes, from the most to the least
  important, is built using a 2000-tree random forest with the full
  attribute set.

\item The most important attribute is selected and all other
  attributes with a Spearman correlation coefficient above 80\% with
  this one are discarded.

\item A new ranked attribute list is built re-running a random forest
  with the selected and the remaining attributes.

\item The second most important attribute is selected and all other
  attributes highly correlated with any of the first two are
  discarded.

\item This process is iterated until a full ranked list of
  not-too-correlated attributes is obtained.
\end{enumerate}

\begin{figure}
 \includegraphics[width=\columnwidth]{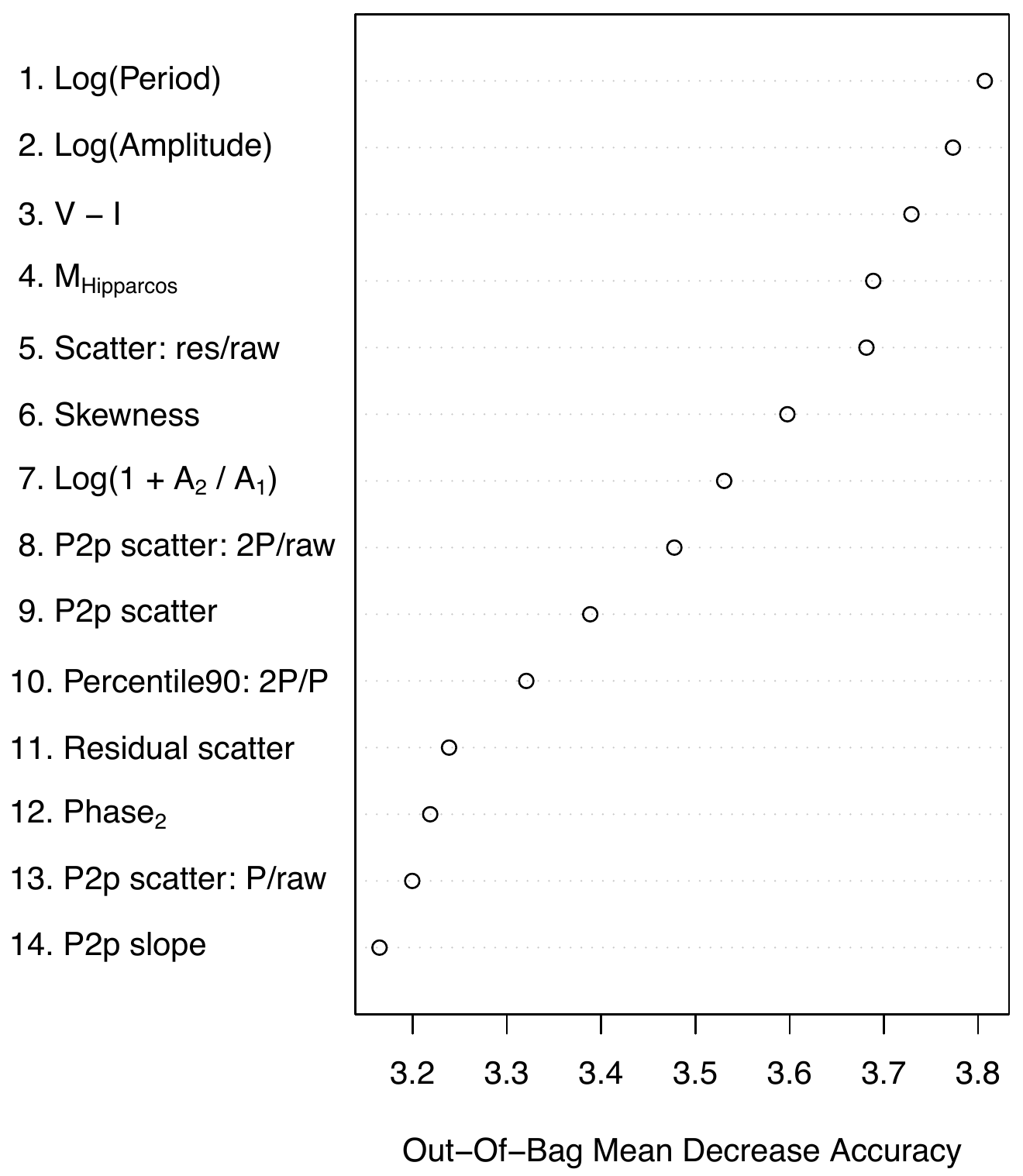}
 \caption{The ranked list of the 14 most important, not-too-correlated
   attributes (defined in Sect.~\ref{SECT_ATTR_DESCR}). The Spearman
   correlation coefficient of any of the above attribute pairs is
   smaller than 80\%. The attribute importance is measured with the
   random forest Out-Of-Bag (OOB) mean decrease accuracy. }
\label{FIG_OOB_ACC}
\end{figure}

This procedure is somewhat unstable if the number of attributes is too
large. The most important attributes are always highly ranked, but the
order of the moderately important ones may change drastically from one
run to the next. Clearly, the importance measurement of a given
attribute depends to some level on the background of the other
attributes. This is even amplified if the attributes are
correlated. If the attribute under evaluation is highly correlated
with other ones, replacing that attribute with random noise does not
affect much the results as the other attributes have similar
classification power. An effective way around this difficulty is to
remove a large fraction of the least important attributes before
starting the above recursive procedure.

Some astronomical insight is also injected into this selection
process. When two, or several, attributes have similar importance, the
one with a simpler and/or more widely used definition is preferred.
Figure~\ref{FIG_OOB_ACC} displays the results of the above attribute
ranking procedure for the 14 most important attributes. A detailed
attribute description is provided below in Sect.~\ref{SECT_ATTR_DESCR}.

\subsection{Towards a minimum attribute list \label{SECT_ANDY}}

\begin{figure}
 \includegraphics[width=\columnwidth]{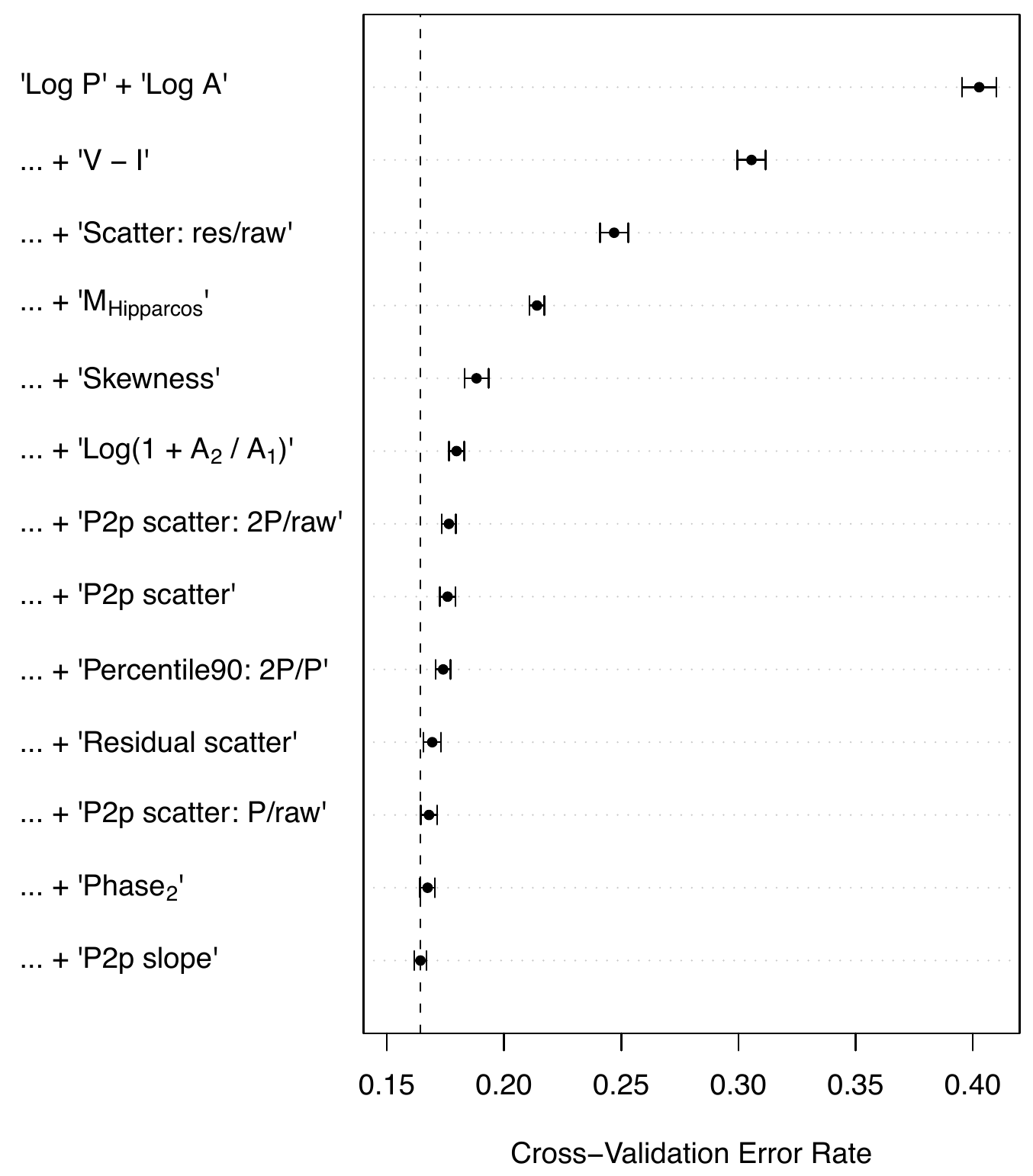}
 \caption{Evolution of the Cross-Validation error rate as more and
   more attributes are added into the classification process. The
   seven most important attributes already drive the error rate under
   18\%, while a minimum of 16.4\% is reached with an additional seven
   attributes. As explained in the main text, some randomness is
   included in our classification process. As a consequence, the
   attribute order can vary slightly in the different CV
   experiments. The attribute name provided at a given line in this
   figure is the name of the attribute appearing most frequently at
   that position in the different experiments.}
\label{FIG_CV_ERR}
\end{figure}

The procedure described in the last section is used to derive a ranked
list of not-too-correlated attributes. The importance value decreases
in the list but it never reaches zero. A key question is where to cut
the list. Are all attributes really useful? Or, are the low-importance
attribute contributions already included in those of more important
attributes? This second possibility is more likely as many of the
low-importance attributes are correlated at some level with some of
the more important attributes in the list.

In order to reduce the number of attributes, a variant of the method
proposed by \citet{svetnik} is adopted. Only the list of
not-too-correlated attributes, derived as described in the previous
section, is used through the following algorithm.

\begin{enumerate}[1.]
\item The data is partitioned for a 10-fold cross-validation (CV).

\item On each CV training set, a ranked list of attributes is
  established using the random forest importance measures as described
  in the previous section.

\item On each CV training set, a model is trained on all attributes
  and used to predict types for the CV test set. The CV error rate is
  recorded and the process is repeated after removing the least
  important attribute. Iterating by removing one attribute at a time
  and stopping when only 2 attributes are left, a vector of CV error
  rates is obtained for an attribute number ranging from 2 to the
  total.

\item At the end of the 10-fold process, a mean error vector is
  computed by taking the mean of the 10 values obtained for each
  attribute sub-set.

\item Steps 1 to 4 are repeated 20 times. The mean value and the
  standard deviation of the 20 CV mean errors are computed for each
  attribute number, combining the results of the classification
  experiments achieved with a specific attribute number.
\end{enumerate}

Figure~\ref{FIG_CV_ERR} shows the error rates as a function of the
number of attributes resulting from the above procedure. The optimum
number of attributes can then be inferred from this figure. As the
attribute number increases, Fig.~\ref{FIG_CV_ERR} shows that the error
rates first decrease and then level-off at some value. A
cross-validation error rate under 18\% is obtained with the first
seven most important attributes and a minimum of 16.4\% is reached
with seven more attributes. The large number of additional attributes
tested in this study are not mentioned as they do not lead to any
further improvement of the classification results.

It is interesting to note that Figure~\ref{FIG_CV_ERR} is much more
contrasted than Fig.~\ref{FIG_OOB_ACC}. While the latter indicates a
steady, almost linear decrease in attribute importance, the drop in
error classification, and hence in attribute merit, seen in
Fig.~\ref{FIG_CV_ERR} is much more abrupt. This is probably due to the
fact that the importance displayed in Fig.~\ref{FIG_OOB_ACC} is
measured against the background of the other 13 attributes. As there
is some remaining correlation between attributes, some of the other
attributes can compensate for the loss of the specific, evaluated
attribute, whose values have been permuted (see
Sect.~\ref{SECT_ATTR_IMP} and \ref{SECT_ATTR_CORR}).

\subsection{The most important attributes \label{SECT_ATTR_DESCR}}

\begin{figure*}
 \includegraphics[width=16cm]{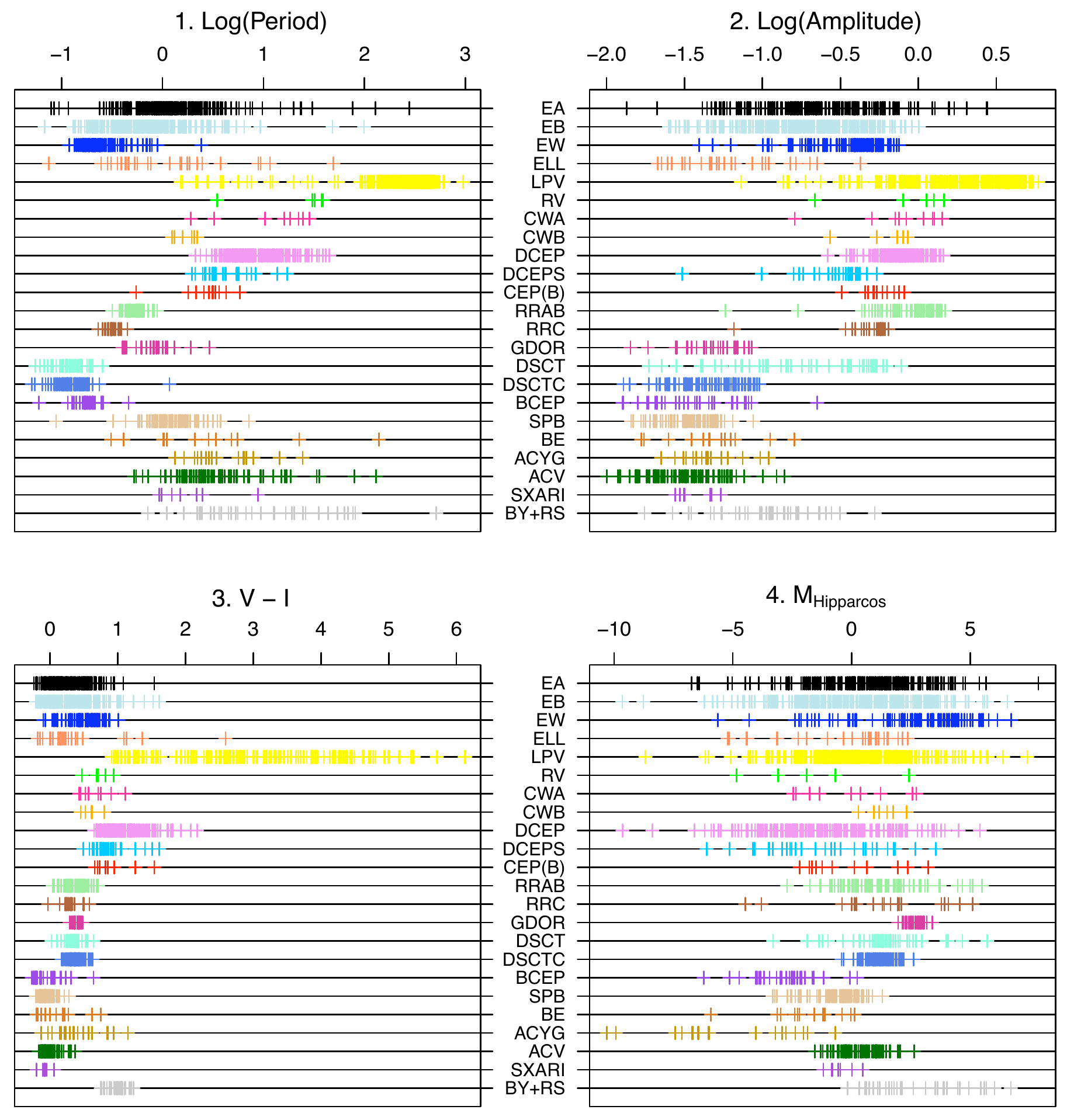}
 \caption{Distributions of the four most important attributes obtained for the training-set members.}
\label{FIG_ATTR14}
\end{figure*}

\begin{figure*}
 \includegraphics[width=16cm]{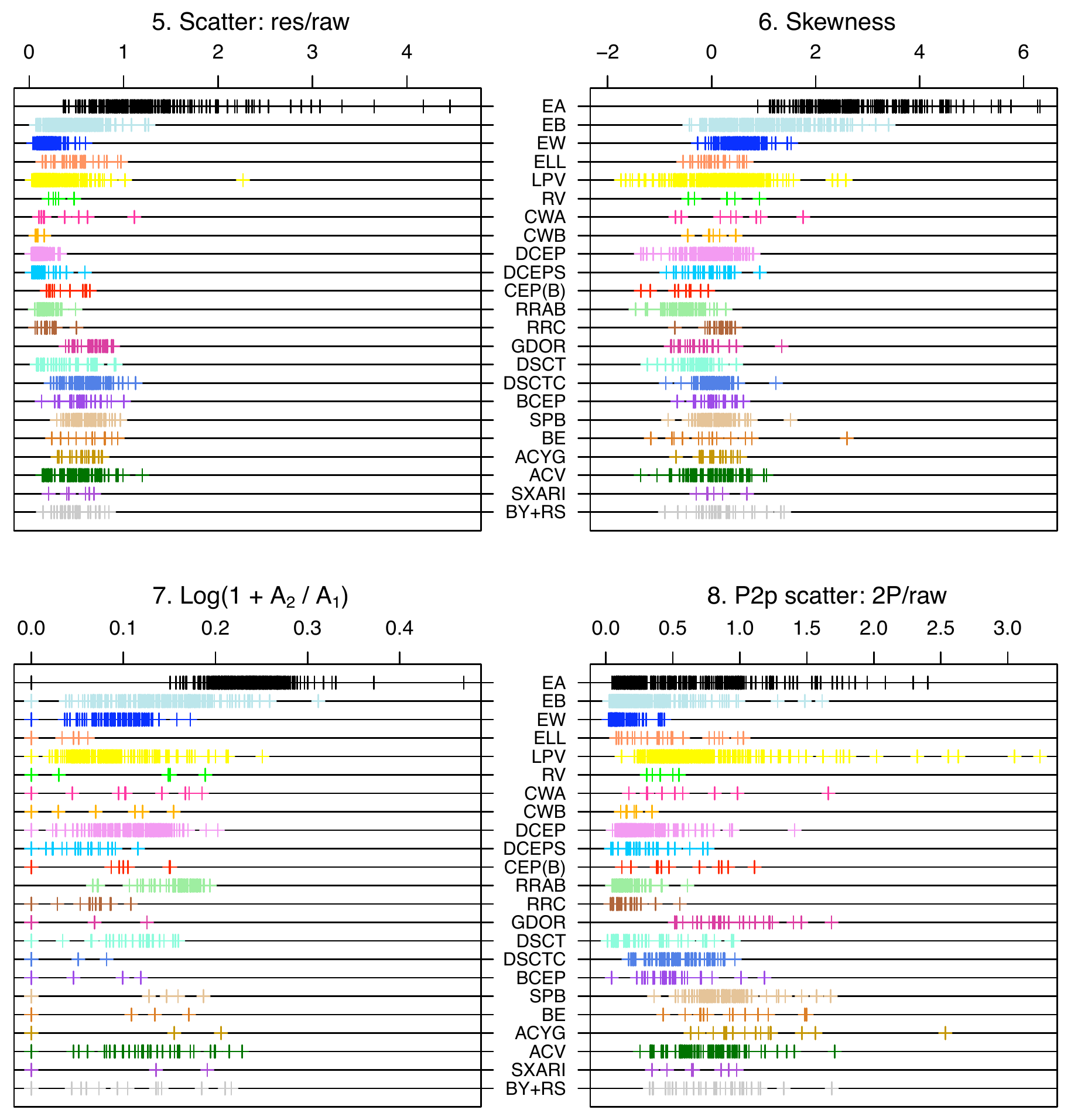}
 \caption{Distributions of the attributes ranking 5 to 8 in random
   forest importance obtained for the training-set members.}
\label{FIG_ATTR59}
\end{figure*}

The 14 most important attributes listed in Fig.~\ref{FIG_OOB_ACC} and
\ref{FIG_CV_ERR} are defined below.

\begin{enumerate}[1.] 
\item {\bf Log(Period)} : decadic log of the period extracted with
  the Lomb-Scargle method (see Sect.~\ref{SECT_PER}).

\item {\bf Log(Amplitude)} : decadic log of the amplitude of the
  light-curve model.

\item {\bf V - I} : the mean V-I colour.

\item {\bf M$_{\rm Hipparcos}$} : a \hip absolute magnitude derived
  from the parallaxes neglecting interstellar absorption. Because of
  measurement uncertainties, some stars have negative parallax
  values. Each of these values is replaced by a positive value taken
  randomly from a Gaussian distribution with zero mean and a standard
  deviation equal to the measurement uncertainty. In many cases, the
  derived absolute magnitudes represent lower limits as the parallax
  measurements are not significant.

\item {\bf Scatter: res/raw} : Median Absolute Deviation (MAD) of the
  residuals (obtained by subtracting model values from the raw light
  curve) divided by the MAD of the raw light-curve values around the
  median.

\item {\bf Skewness} : unbiased weighted skewness of the magnitude
  distribution.

\item{\bf  Log(1 + A$_{\rm 2}$ /  A$_{\rm 1}$)} : decadic log of the
  amplitude ratio between the second harmonic and the fundamental (plus one,
  to avoid negative values).

\item {\bf P2p scatter: 2P/raw} : sum of the squares of the magnitude
  differences between pairs of successive data points in the light
  curve folded around twice the period divided by the same quantity
  derived from the raw light curve.

\item {\bf P2p scatter} : median of the absolute values of the
  differences between successive magnitudes in the raw light curve
  normalized by the Median Absolute Deviation (MAD) around the median.

\item {\bf Percentile90: 2P/P} : the 90-th percentile of the absolute residual
  values around the 2P model divided by the same quantity for the
  residuals around the P model. The 2P model is a model recomputed
  using twice the period value.

\item {\bf Residual scatter} : mean of the squared residuals around
  the model.

\item {\bf Phase$_{\rm 2}$} : phase of the second harmonic after
  setting the phase of the fundamental to zero by an appropriate
  transformation Phase$_{\rm 2} = \arctan(\sin(\varphi_{2} - 2
  \varphi_{1} ), \cos(\varphi_{2} - 2 \varphi_{1} ))$
  \citep{Debosscher+2007}.

\item {\bf P2p scatter: P/raw} :  median of the absolute values of the
  differences between successive magnitudes in the folded light curve
  normalized by the Median Absolute Deviation (MAD) around the median
  of the raw light curve.

\item {\bf P2p slope} : sum of the square of the slopes of lines
  joining the data points before and after a number of selected
  outliers towards faint magnitude (e.g., data points during
  eclipses). This is set to zero if there are no such outliers in the
  light curve.
\begin{equation}
\mbox{P2p slope}=\sum_i \left[\left(\frac{d_i-d_{i-1}}{t_i-t_{i-1}} \right)^2 + \left( \frac{d_{i+1}-d_i}{t_{i+1}-t_i}\right)^2\right]
\,\,\, \mbox{for }d_i>3, \nonumber
\end{equation}
with $d_i=(y_i-P_{25})/\delta_i$, $\delta_i=(\sigma_i^2+\delta^2)^{1/2}$, and $\delta=P_{25}-P_5$, where $y_i$ and $\sigma_i$ are the observed magnitude and its error, respectively, at time $t_i$, and $P_n$ is the $n$-th percentile of the magnitude distribution.

\end{enumerate}

\subsection{Attribute display}

Figures \ref{FIG_ATTR14} and \ref{FIG_ATTR59} display the
distributions of the 8 most important attributes for each of the
variability types.

\subsection{Classification error analysis \label{SECT_ERR_ANA}}

\begin{figure*}
 \includegraphics[width=13cm]{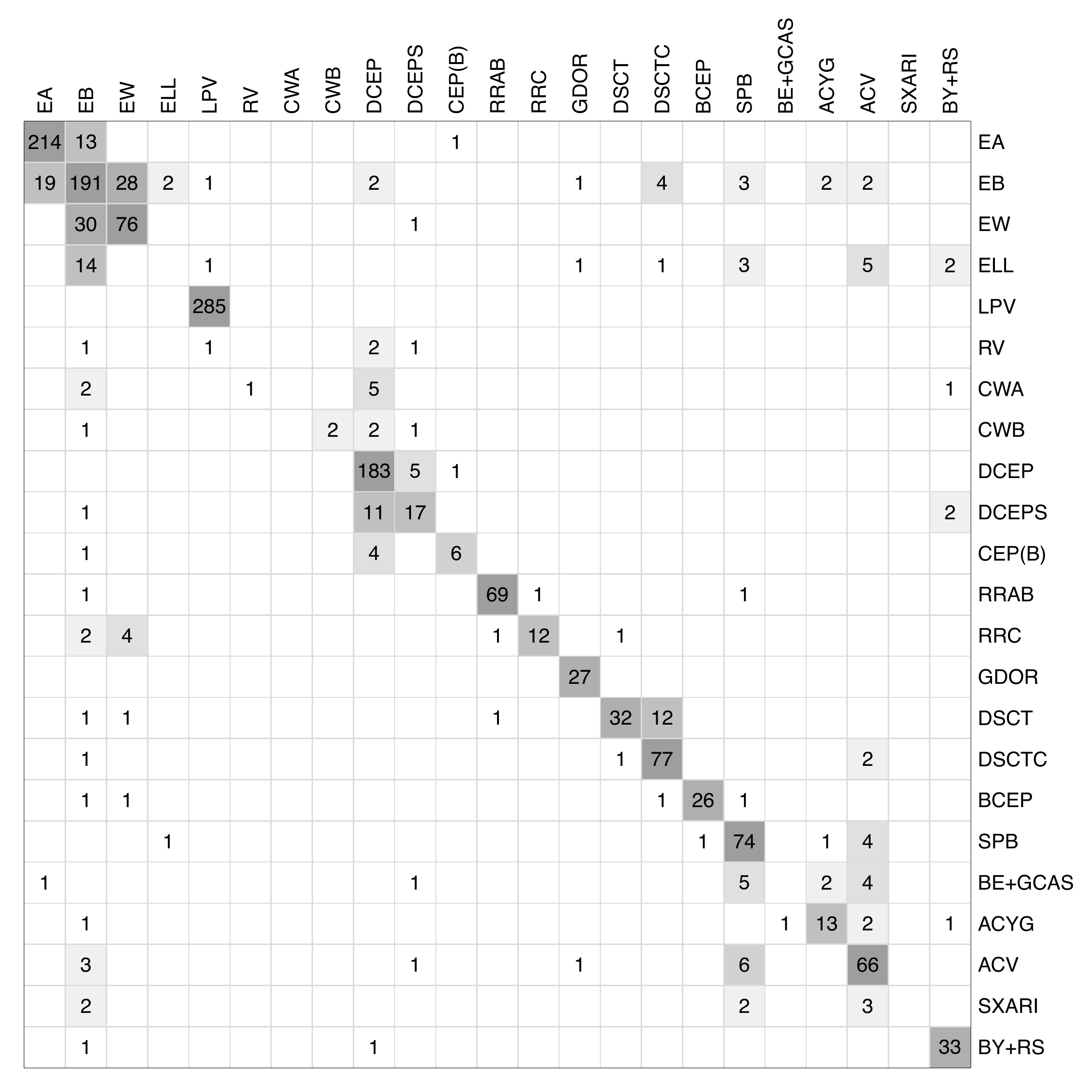}
 \caption{Confusion matrix obtained for our training set with a
   10,000-tree random forest classification using a group of 14
   attributes. The rows indicate the reference types resulting from
   the literature survey (see Sect.~\ref{SECT_TS}, while the columns
   represent the classifier {\em predicted} types. Type labels are as
   described in Table~\ref{TAB_TS}.}
\label{FIG_CM}
\end{figure*}

Figure~\ref{FIG_CM} displays the confusion matrix resulting from a
10,000-tree random forest classification of our training-set
stars. The 14 attributes described in Sect.~\ref{SECT_ATTR_DESCR} are
used and the number \mtry (see Sect.~\ref{SECT_ALGO}) of attributes
tried at each node is three. The matrix rows indicate the reference
types resulting from the literature survey presented in
Sect.~\ref{SECT_TS}, while the columns represent the classifier {\em
  predicted} types. The classification process fails to separate some
of the types, namely (1) \dsct from \sxphe, (2) \be from \gcas, and
(3) \by from \rs.  These types are pairwise merged in
Fig.~\ref{FIG_CM} to improve the matrix readability.  In principle,
the \sxphe could be separated if a metallicity estimator was used in
the classification and \by and \rs could be distinguished if better
absolute magnitudes were available. The case of the \be and \gcas is
discussed below.

The overall classification error rate derived from the Out-Of-Bag
samples is 15.7\%. It is slightly lower than the 16.4\% presented in
Fig.~\ref{FIG_CV_ERR} because of the above-mentioned merging of six
types. However, this overall rate does not bear much meaning as
confusions within groups of similar stars are less problematic than
others. The most important confusion cases are detailed in the following
sub-sections. It is important to remember that random forest involves
randomness in the sample bootstrapping and node attribute selection
(see Sect.~\ref{SECT_ALGO}).  As a consequence, the confusion matrices
obtained in successive identical runs differ slightly at the level of
a few cases.

\subsubsection{Eclipsing binaries and ellipsoidal variables}

As already alluded to in Sect.~\ref{SECT_PER_ECL}, eclipsing binaries
are expected to be a difficult case. It is therefore not surprising to
see from Fig.~\ref{FIG_CM} that they are involved in the most
important confusion cases. The classification disperses 17 \eb into
other types of non-eclipsing variables. There are 14 \elli variables
mis-classified as \eb while 13 of them are scatter into 6 other types.
In addition, 19 and 6 cases of diverse other non-eclipsing and
non-ellipsoidal variables are unduly classify as \eb and \ew,
respectively.

Could this confusion be diminished if the full training set is first
separated into eclipsing and non-eclipsing variables? To investigate
this issue the attribute ranking and selection procedure is repeated
considering two type groups, \ea, \eb, \ew and \elli on the one side,
and all other types on the other. The resulting attribute ranking is
quite different but the classification results do not improve. It is
possible to lower the number of variables falsely classified as
eclipsing binaries to about 20 cases, but then, the number of
mis-classified eclipsing binaries increases to about 50, so that the
total is slightly worse than the result of a direct classification
into all types.

The trouble is that the light curves of some \eb, \ew and \elli are
quite symmetrical and resemble those of other variability types. In
addition, stellar properties such as colour and absolute magnitude can
take almost any possible value as they are the combination of the
properties of the two stars of the binary system. As a consequence,
it is likely that these confusion cases represent a true physical
difficulty that cannot be fully solved by any classification method.

\subsubsection{Cepheid and Cepheid-like variables}

As can be clearly seen in Fig.~\ref{FIG_CM}, all types of Cepheid-like
variables are confused with the Delta Cepheid type. More precisely the
following cases can be listed.

\begin{enumerate}[1.] 

\item The \cw (\cwa and \cwb) variables are population II Cepheid
  stars of lower absolute magnitude (and smaller mass). The Hipparcos
  parallax measurements are not significant for these relatively
  bright and remote stars. As a consequence, the derived absolute
  magnitudes are dominated by noise and this explains the confusion as
  these stars are otherwise similar to \dcep. The \cw and \dcep
  variables could be separated using a metallicity indicator or more
  reliable luminosity estimates.

\item The \dceps stars are Cepheids with smaller amplitude and period
  values, which probably pulsate in the first overtone. There is
  however an overlap with the \dcep stars in the log(Period) -
  log(Amplitude) diagram and this probably explains the confusion
  cases. Out of the sample of 31 \dceps stars, 17 are correctly classified
  while 11 fall wrongly in the \dcep category.
 
\item The \cepb stars are Cepheids which exhibit two or more pulsation
  modes. They could almost certainly be better singled out by
  searching for additional significant periods and using them in the
  characterisation and classification processes. This concerns however
  a small number of stars (only 4 out of 10 \cepb stars are wrongly
  classified as \dcep) and it is outside of the scope of this paper,
  which is restricted to single-period analyses.

\end{enumerate}

In addition, there seems to exist a not well understood confusion
between \rv and \dcep although small number statistics here is a
limitation.

\subsubsection{Blue variables}

The third case of confusion concerns the blue variables.  First \be
and \gcas have been put together in Fig.~\ref{FIG_CM}.  These types
can only be separated on the basis of long term behaviour. \gcas show
eruptive, non-periodic events
\citep{Samus2009}\footnote{http://www.sai.msu.su/gcvs/gcvs/iii/vartype.txt}
and our sample includes only those stars where the observed signal is
periodic. The short term periodic behaviour of some \be + \gcas is also
similar to the one observed in the \spb stars
\citep[e.g.,][]{Diago2009}. As a consequence, it is not surprising to
observe a confusion between \spb and \be + \gcas.

The confusion between \be+ \gcas and \acv cannot be so easily
understood. It most probably comes from a true confusion between \acv
and \spb, seen at the ~10\% level in Fig.~\ref{FIG_CM} , which also
concerns the few \sxari which are physically relatively close in
attribute space to the \acv.

\subsection{Random forest and linear discriminant analysis}

The Linear Discriminant Analysis (LDA) \citep{lda_0, Hastie+09}
approach is applied to derive an optimum set of independent Linear
Discriminants (LDs). The goal is to run a random forest classification
using these LDs as an alternative set of attributes. 

In the attribute multi-dimensional space, objects of a particular type
can be visualized as a ``cloud'' of data points. The complete training
set is then viewed as a set of generally overlapping clouds that are
to be separated by the process of classification. The idea of LDA is
to derive linear transformations of the attributes which maximise the
ratio of the cloud centre variance divided by the variance of the data
points within the clouds. In other words, the transformation seeks to
rotate the axes so that when the objects are projected on the new
axes, the differences between the different clouds (i.e., types) are
maximised.

The LDA-based classification scheme goes through the following steps.

\begin{enumerate}[1.] 

\item For each attribute, the distribution of values is standardised by
  subtracting the mean and dividing by the standard deviation.

\item A LDA is carried out. The resulting LDs are ranked as a function
  of the singular values, i.e., the most important LD is the one with
  the largest ratio of the variance of the group centres over the
  within-group variance.

\item The attributes are ranked according to their maximum
  contribution to any of the most important LDs.

\item Different attribute selection schemes are used, iterating and
  removing one, or a few of the least important attributes and of the
  highly correlated ones in each of the successive steps. Although,
  this process is completely independent from the one used previously
  for random forest, the final list of selected attributes is very
  similar, with period, amplitude and colour attributes always
  standing out as the best three.

\item The resulting list of attributes is used in a final LDA.  The
  LDs are computed for all stars and used for a random forest
  classification.

\item The classification errors are estimated using the OOB sample and
  a 10-fold cross-validation method. 
\end{enumerate}

Although many attempts have been performed varying the attribute
selection scheme, the resulting classification errors are always
significantly worse (by at least 3\%) than those obtained when
applying random forest to the original attributes. The derived LDs
are less correlated than the original attributes (even when
relaxing the selection criteria and accepting more highly correlated
attributes), but surprisingly, it did not lead to better results in
our case. Thus, LDA is not used to produce any of the results
presented in this paper.

\section{Degradations due to errors in the period determination \label{SECT_PERR}}

As shown in Fig.~\ref{FIG_PER_OTHER}, in some cases, the period values
resulting from the search done in this work are completely different
from the \hipcat values. Although, the latter values are probably more
reliable as they were visually checked, our classification is based on
our own values as the idea is to evaluate the performance of an
automated classification process (see Sect.~\ref{SECT_PER}). It is
however interesting to investigate the classification degradation
induced by wrong period values. The stars with incorrect periods can
be traced to evaluate how well they are classified. The level of
confusion observed for these stars can be compared with that seen for
stars with correct periods.

We exclude from this comparison the eclipsing binaries and the
ellipsoidal variables as it is known (see Sect.~\ref{SECT_PER_ECL})
that the period found is systematically half that of the true period
for these stars. Since the eclipsing binary periods span a wide range
of values, the confusion due to an incorrect period is likely to be
less severe than that observed for other stars.

\begin{table}
\begin{center}
  \caption{\label{TAB_WRONG_P}  Confusion induced by incorrect periods for
    non-eclipsing variables}
  \begin{tabular}{lrrr}
  \hline
    &     \multicolumn{1}{c}{Total}    & Misclassified stars & Error Rate   \\
 \hline
Stars with correct period       &    951     &   102    &  10.7\% \\
Stars with incorrect period         &      93     &     20    &   21.5\%     \\
All stars                             &  1044     &    122   &   11.7\% \\
\hline 
\end{tabular} \\
\end{center}
\end{table}

The comparison presented in Sect.~\ref{SECT_PER_OTHER} shows that out
of the total of 1044 non-eclipsing variables, an incorrect period is
found for 93 stars. Table~\ref{TAB_WRONG_P} displays the
classification errors obtained in the different cases. The eclipsing
binaries are excluded, but the cases where a non-eclipsing is classified
as an eclipsing (or an ellipsoidal) variable are accounted for in the 
numbers of misclassified stars.

Although statistical uncertainties due to the small numbers is a
limitation, this table shows that 73 out of the total of 93 stars with
an incorrect period are successfully classified into their proper
types. This is surprising as the period always stands as the most
important attribute. Somehow, other attributes, such as the amplitude,
the colour, or the pseudo absolute magnitude, compensate and safeguard
against an incorrect classification in an important number of cases.
 
%\section{Impact of a possible reddening}

\section{Comparison with a multi-stage classifier \label{SECT_MULT}}

\begin{figure}
   \centering
   \includegraphics[width=\columnwidth]{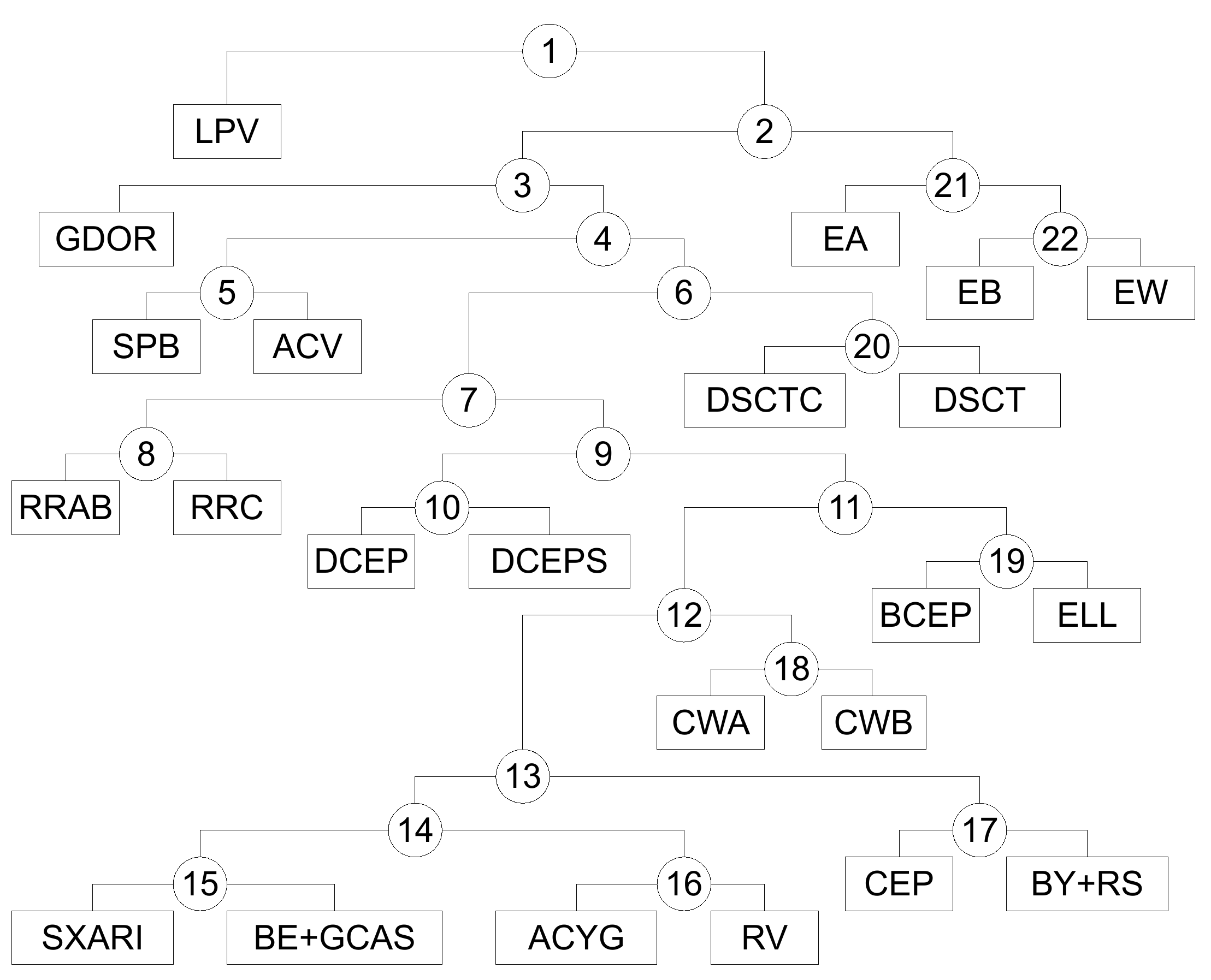}
   \caption{Sequential scheme for the separation of all variability
     types with dichotomic classifiers.}
   \label{ms-tree}
\end{figure}

For comparison with the results shown in Sect.~\ref{SECT_RF}, a
methodology based on a divide-and-conquer approach is applied whereby
the overall classification problem with 26 variability types included
in Table~\ref{TAB_TS} is sequenced into several stages. The
variability zoo in that table is grouped into categories and
subcategories, and the classification of a star proceeds by assigning
a probability vector for each category and subcategory until one of
the variability types defined in Table \ref{TAB_TS} is reached (the
leaves of the tree defined in Fig.~\ref{ms-tree}). This multi-stage
scheme is defined in more detail in \citep{Blomme+2011}. Here a
different grouping of variability types based on an automatic
definition of categories is tested.

The algorithm used to define the multi-stage scheme shown in Figure
\ref{ms-tree} is based on the confusion matrix obtained by using a
monolithic, single-stage classifier. From this, the similarity between
types is determined according to the metrics

\begin{equation}
{\rm Similarity}(\mathcal{T}_i,\mathcal{T}_j)=
\begin{cases}
1, & i=j \\
\frac{X_{ij}+X_{ji}}{X_{ij}+X_{ji}+X_{ii}+X_{jj}}, & i\ne
j\\
\end{cases}
\end{equation} 

\noindent where $\mathcal{T}_i$ represents type $i$ and $X_{ij}$ represents the
element in the $i$-th row and $j$-th column of the confusion matrix.

The idea behind this metric is that types that are easy to separate
should be in the topmost levels of the scheme since they do not affect
other types too much, and the most problematic types are positioned at
the bottom of the sequence. The algorithm starts from the complete set
of types and, in each step, the two most similar types are merged into
a new type, and the similarities are calculated again over the new set
of types. Thus, the final multi-stage tree is composed of a series of
dichotomic classifiers.

The advantage of the divide-and-conquer approach is that the optimal
set of attributes used for classification and the optimal
classification algorithm can be selected in each node of the tree. In
our case, the best algorithm in each node is selected from a set
composed of Bayesian Networks \citep{Pearl1988} and Gaussian Mixtures
\citep{Debosscher+2007}. These two methods are chosen because both
Bayesian Networks and Gaussian Mixtures allow for a simple procedure
in order to account for missing attributes. In both cases this is
accomplished via marginalisation of the posterior type probabilities
given the attributes. In the case of the Gaussian Mixture classifier
this can be achieved analytically as

\begin{equation}
p(\mathcal{T}|\vec{x}_{\rm avail}) = 
\int p(\mathcal{T}|\vec{x}_{\rm avail},\vec{x}_{\rm missing}) 
\, p(\vec{x}_{\rm missing})\, {\rm d}\vec{x}_{\rm missing},
\end{equation}

\noindent where $\vec{x}_{\rm avail}$ and $\vec{x}_{\rm missing}$ are
the subsets of available and missing attributes respectively, which
make up the complete attribute vector $\vec{x}$, and the probability
density functions are always normal. $p(\mathcal{T}|\vec{x}_{\rm
  avail},\vec{x}_{\rm missing})$ is an outcome of the training-set
classification and the distribution of the missing attribute
$p(\vec{x}_{\rm missing})$ can be established from the sub-sample of
other stars for which $\vec{x}_{\rm missing}$ is available or from
astrophysical knowledge of the distribution.

In each node, the classifier that shows the smallest misclassification
rate is chosen. The misclassification rate is obtained by averaging
the misclassification rates obtained in 10 experiments of 10-fold
cross validation (the so-called multiple runs k-fold cross validation
\citep{Bouckaert2003}; see below for details). This type of experiment
allows for the comparison between two classifiers by statistically
testing the null hypothesis that the two classifiers perform equally
well. Here for simplicity Bayesian Networks is selected in those nodes
where the null hypothesis could not be rejected (i.e., where there was
not sufficient evidence that one of the classifiers outperformed the
other).

As stated above, one of the advantages of the multi-stage
classification is that it allows for context dependent feature
selection. That is, the optimal attribute set for classification can
be selected in each node of the tree. This is particularly useful for
variability classification where the variables that discriminate
between types depend on the types themselves. The procedure adopted
here for variable selection starts with an empty set of attributes in
each node. Then, the attribute that conveys the largest mutual
information with the type is added. Attributes are added to the set
following this greedy strategy until the addition of a new attribute
produces an increase in the mutual information of less than 0.1. This
threshold is found to avoid in most of the cases the inclusion of
attributes which are deemed irrelevant on the basis of expert
astronomical knowledge. These irrelevant attributes are sometimes
picked by the algorithm due to spurious correlations caused by the
small training set sample sizes.

\begin{table*}
 % \begin{center}
   \caption{\label{TAB_TRAINING} A sample of the \hip training set star list with
     literature types and attribute values. The full table is
     available at (online link)}
   \begin{tabular}{@{}rlrrrrrrrrrrrrrr@{}}
   \hline
Hip
& Type
& \multicolumn{1}{c}{\rotatebox{90}{Log(Period)}}
& \multicolumn{1}{c}{\rotatebox{90}{Log(Amplitude)}}
& \multicolumn{1}{c}{\rotatebox{90}{V - I}}
& \multicolumn{1}{c}{\rotatebox{90}{M$_{\rm Hipparcos}$}}
& \multicolumn{1}{c}{\rotatebox{90}{Scatter: res/raw}}
& \multicolumn{1}{c}{\rotatebox{90}{Skewness}}
& \multicolumn{1}{c}{\rotatebox{90}{Log(1 + A$_{\rm 2}$ /  A$_{\rm 1}$)}}
& \multicolumn{1}{c}{\rotatebox{90}{P2p scatter: 2P/raw}}
& \multicolumn{1}{c}{\rotatebox{90}{P2p scatter}}
& \multicolumn{1}{c}{\rotatebox{90}{Percentile90: 2P/P}}
& \multicolumn{1}{c}{\rotatebox{90}{Residual scatter (\%)}}
& \multicolumn{1}{c}{\rotatebox{90}{P2p scatter: P/raw}}
& \multicolumn{1}{c}{\rotatebox{90}{Phase$_{\rm 2}$}}
& \multicolumn{1}{c}{\rotatebox{90}{P2p slope}} \\
 \hline
     8 & LPV &  2.5229 &  0.61 &  3.92 &   2.12 &  2.60 &  0.61 & 0.00 & 0.48 & 0.07 & 0.68 & 11.77 &  1.00 &  1.57 & 1.63 \\
    63 & ACV &  0.5726 & -1.40 & -0.03 &  -0.24 &  1.62 & -0.62 & 0.13 & 0.85 & 0.75 & 1.10 &  0.01 &  1.20 &  1.70 & 0.00 \\
   109 & DSCTC & -0.7819 & -1.41 &  0.45 &   0.95 &  0.95 & -0.18 & 0.00 & 0.84 & 1.36 & 1.01 &  0.02 &  1.13 &  1.57 & 0.00 \\
   226 & RRAB & -0.3068 &  0.11 &  0.29 &  -0.47 & 11.16 & -0.75 & 0.17 & 0.11 & 0.51 & 0.95 &  0.11 &  0.30 &  2.32 & 0.00 \\
   270 & EA & -0.1576 & -0.71 &  0.16 &   0.77 &  0.68 &  2.00 & 0.17 & 1.09 & 1.31 & 1.01 &  0.55 &  1.09 & -1.51 & 4.15 \\
   316 & DSCTC & -0.7693 & -1.19 &  0.42 &   0.99 &  1.73 &  0.01 & 0.00 & 0.49 & 0.99 & 0.96 &  0.04 &  0.69 &  1.57 & 2.37 \\
   344 & LPV &  2.5100 &  0.71 &  3.91 &   4.35 &  7.38 & -0.20 & 0.00 & 1.62 & 0.06 & 0.60 &  6.25 &  0.99 &  1.57 & 2.59 \\
   623 & GDOR & -0.0375 & -1.38 &  0.44 &   3.40 &  1.12 & -0.36 & 0.00 & 1.40 & 0.94 & 1.01 &  0.08 &  1.42 &  1.57 & 0.00 \\
   703 & LPV &  2.5591 &  0.34 &  1.53 &   1.00 &  5.19 &  0.47 & 0.04 & 0.71 & 0.15 & 1.15 &  1.75 &  1.02 & -0.55 & 2.78 \\
   746 & DSCTC & -0.9955 & -1.49 &  0.40 &   1.24 &  3.05 &  0.32 & 0.00 & 0.21 & 1.59 & 0.95 &  0.00 &  0.32 &  1.57 & 3.46 \\
\hline
 \end{tabular}
% \end{center}
\end{table*}

\begin{table*}
 % \begin{center}
  \caption{\label{TAB_TEST} Results obtained for the \hip stars
    excluded from the training set. This table shows the \hip numbers,
    the literature types, the predicted types and the attribute
    values for a subset of the sample. The full table is available at
    (online link)}
   \begin{tabular}{@{}rllrrrrrrrrrrrrrr@{}}
   \hline
Hip
& Type
& \multicolumn{1}{c}{\rotatebox{90}{Predicted Type}}
& \multicolumn{1}{c}{\rotatebox{90}{Log(Period)}}
& \multicolumn{1}{c}{\rotatebox{90}{Log(Amplitude)}}
& \multicolumn{1}{c}{\rotatebox{90}{V - I}}
& \multicolumn{1}{c}{\rotatebox{90}{M$_{\rm Hipparcos}$}}
& \multicolumn{1}{c}{\rotatebox{90}{Scatter: res/raw}}
& \multicolumn{1}{c}{\rotatebox{90}{Skewness}}
& \multicolumn{1}{c}{\rotatebox{90}{Log(1 + A$_{\rm 2}$ /  A$_{\rm 1}$)}}
& \multicolumn{1}{c}{\rotatebox{90}{P2p scatter: 2P/raw}}
& \multicolumn{1}{c}{\rotatebox{90}{P2p scatter}}
& \multicolumn{1}{c}{\rotatebox{90}{Percentile90: 2P/P}}
& \multicolumn{1}{c}{\rotatebox{90}{Residual scatter (\%)}}
& \multicolumn{1}{c}{\rotatebox{90}{P2p scatter: P/raw}}
& \multicolumn{1}{c}{\rotatebox{90}{Phase$_{\rm 2}$}}
& \multicolumn{1}{c}{\rotatebox{90}{P2p slope}} \\
\hline
   262 & EA: & EA &  0.3518 & -0.04 &  0.49 &  2.31 &  0.69 &  5.03 & 0.23 & 1.06 & 0.82 & 1.86 &  2.79 & 1.73 & -1.56 & 3.75 \\
   516 & LPV: & LPV &  2.1605 &  0.14 &  2.43 & -3.25 &  2.68 & -0.39 & 0.10 & 0.86 & 0.07 & 0.82 &  3.38 & 1.24 &  2.97 & 0.00 \\
   664 & RS+BY: & RS+BY &  1.6836 & -0.76 &  1.33 & -1.12 &  4.96 &  0.26 & 0.00 & 0.34 & 0.16 & 0.72 &  0.02 & 1.03 &  1.57 & 1.93 \\
   723 & LPV: & RS+BY &  2.5575 & -1.31 &  0.89 &  6.19 &  1.26 &  0.02 & 0.00 & 0.90 & 1.10 & 0.95 &  0.05 & 1.05 &  1.57 & 1.71 \\
   864 & LPV: & RS+BY &  1.5759 & -0.71 &  1.72 & -1.17 &  2.75 & -0.34 & 0.08 & 0.77 & 0.29 & 0.61 &  0.12 & 1.10 &  1.54 & 0.00 \\
   871 & EB: & EA &  1.2616 & -0.05 &  0.73 &  0.32 &  1.61 &  2.70 & 0.20 & 0.27 & 0.85 & 1.23 &  0.29 & 0.83 & -1.81 & 2.73 \\
   988 & LPV: & LPV &  1.6542 & -0.52 &  2.67 & -4.32 &  2.43 & -0.70 & 0.00 & 0.67 & 0.31 & 1.02 &  0.10 & 0.91 & -1.37 & 0.00 \\
  1110 & LPV: & LPV &  1.1795 & -0.56 &  2.06 &  0.28 &  3.18 & -0.41 & 0.00 & 0.68 & 0.25 & 1.01 &  0.37 & 1.23 &  1.57 & 0.00 \\
  1263 & EB: & EA &  0.9776 & -0.62 &  0.46 &  2.26 &  0.91 &  3.19 & 0.21 & 0.94 & 1.50 & 0.91 &  0.32 & 0.79 & -1.60 & 3.41 \\
  1378 & ACV: & SPB & -0.0238 & -1.22 & -0.06 & -2.30 &  1.30 &  0.25 & 0.00 & 0.71 & 1.14 & 1.01 &  0.04 & 0.75 &  1.57 & 1.88 \\
\hline
 \end{tabular}
 %\end{center}
\end{table*}

The comparison between the classification strategy described in
Sect.~\ref{SECT_RF} and the multi-stage classifier is done following
the same procedure \citep{Bouckaert2003} used to select the best
classifier in the nodes of the multi-stage tree. Ten experiments of
10-fold cross validation are carried out. For each 10-fold cross
validation experiment, the misclassification rate of the two
alternative classifiers are subtracted. Let $a_{ij}$ denote the
misclassification rate of one of the classifiers in the $i$-th run and
$j$-th fold, and $b_{ij}$ that of the alternative. Then, the
difference $x_{ij}=a_{ij}-b_{ij}$ is calculated and the values of
$x_{ij}$ within the same run are sorted in increasing order. Finally,
the values of $x_{ij}$ in each fold are averaged over the ten
different runs. Thus, this ends up with 10 sorted average
misclassification rates and the corresponding variance estimates, one
for each fold. Then, the $Z$ statistic is computed as follows

\begin{equation}
Z=\frac{m}{\sqrt{\hat{\sigma}^2}\sqrt{df+1}},
\end{equation}

\noindent where $m$ is the mean of the 100 misclassification rates,
$\hat{\sigma}^2$ is the variance averaged over the 10 fold-wise variance
estimates, and $df$ is the number of degrees of freedom. In our case,
the calibration by \cite{Bouckaert2003} (i.e., $df=10$) is used.

It can be shown that the $Z$ statistic follows a $t$ distribution for
two classifiers that perform equally well (the null hypothesis). In
our case, a value of Z=0.516 is obtained. It corresponds to a
$p$-value of 0.31 which is clearly above any reasonable confidence
threshold. Therefore, there is no evidence to reject the null
hypothesis that the two classification strategies (the random forest
and the multi-stage tree) perform equally well.

\section{Automated classification \label{SECT_TYPES}}

A complete set of attributes is available for 2543 stars out of the
total of 2712 \hip periodic variables. A sub-set of 1661 of these
stars, selected following the procedure described in
Sect.~\ref{SECT_TS}, forms the training set used in previous
sections. There are 882 stars left, for which either only an
uncertain type is available from the literature (832), or no type at
all can be found (50). These stars are classified applying the best
model obtained through the random forest processing of the training
set.

\begin{figure*}
 \includegraphics[width=13cm]{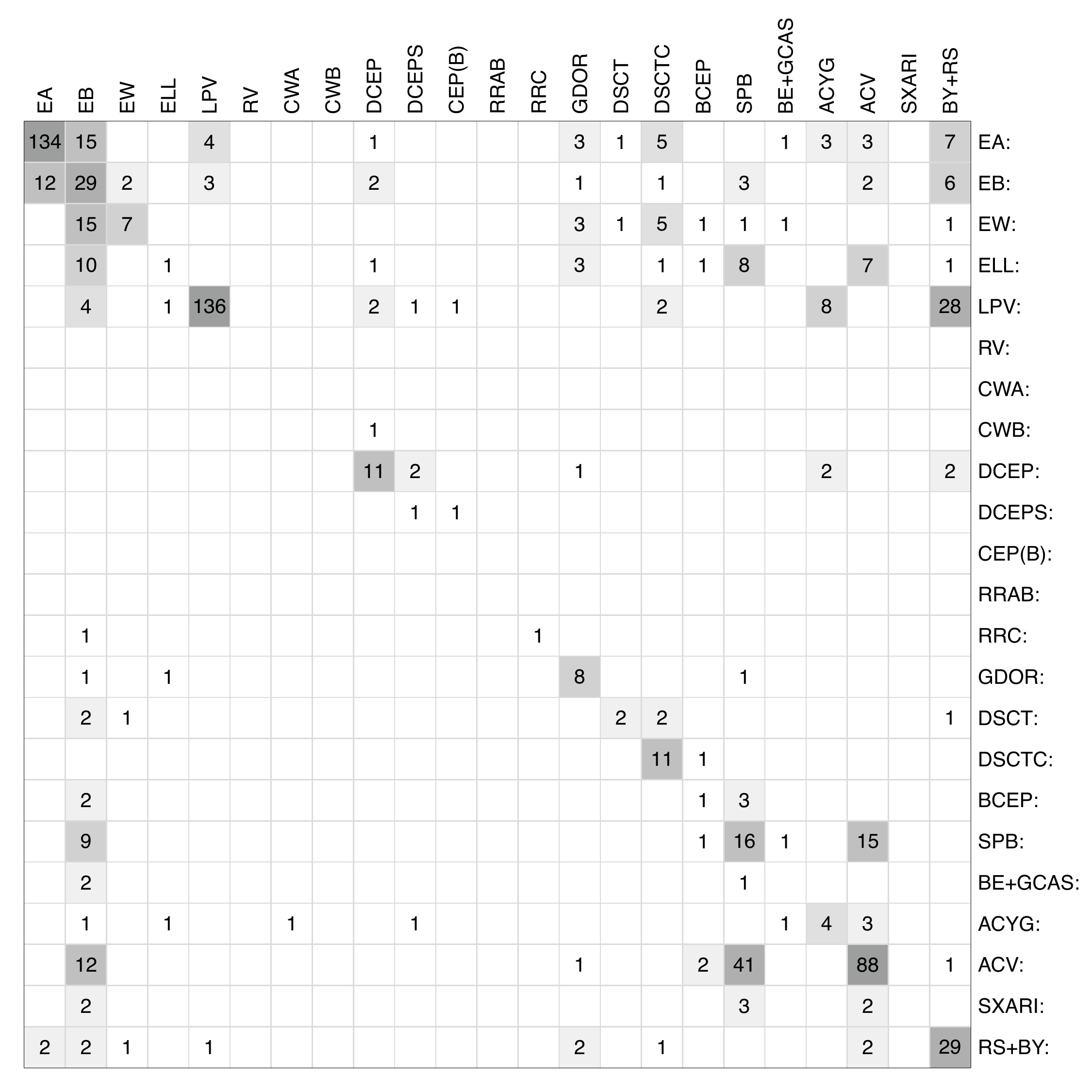}
 \caption{Confusion matrix obtained for a subset of \hip stars with
   uncertain types from the literature. These types are shown in rows
   while columns indicate the types predicted by the random forest model
   derived from the training set analysis. Type labels are
   described in Table~\ref{TAB_TS}.}
\label{FIG_CM_UNCERT_1}
\end{figure*}

\begin{figure*}
 \includegraphics[width=13cm]{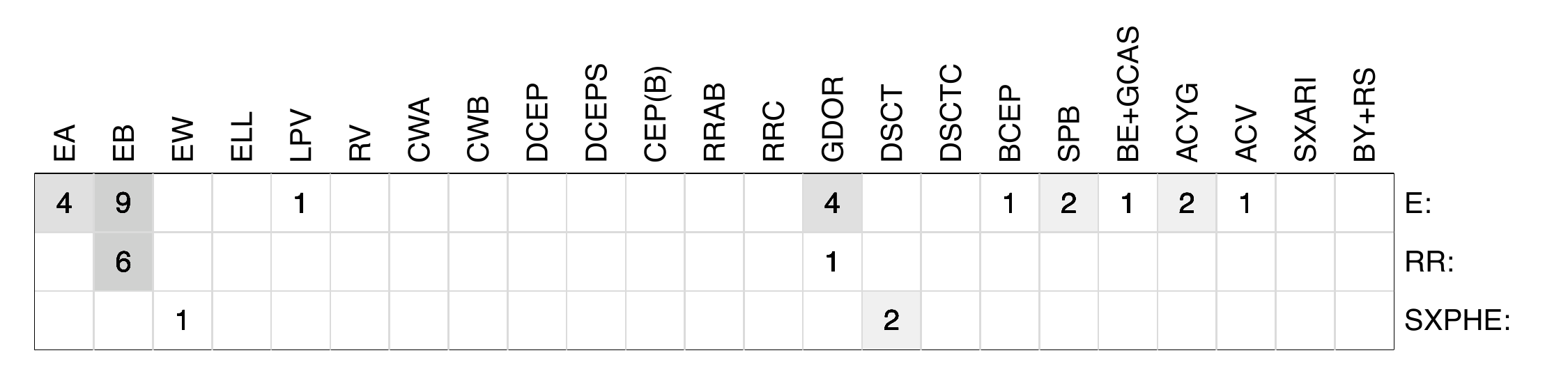}
 \caption{Same as Fig.~\ref{FIG_CM_UNCERT_1} but for stars with types
   that do not directly match any of the types from training set stars.}
\label{FIG_CM_UNCERT_2}
\end{figure*}

Table~\ref{TAB_TRAINING} shows the literature types and the attribute
values for the training-set sample, while Table~\ref{TAB_TEST} shows
the literature types (when available), the predicted types and the
attribute values for the 882 uncertain-type star sample. The predicted
types are also compared to the literature types using a
confusion-matrix type of display in Fig.~\ref{FIG_CM_UNCERT_1} and
Fig.~\ref{FIG_CM_UNCERT_2}. Quite expectedly, the confusion is larger
in these matrices, but the main confusion cases are the same as
those observed with the training set (see Fig.~\ref{FIG_CM} and the
discussion of Sect.~\ref{SECT_ERR_ANA}). Note that only the sub-set of
stars with available uncertain types from the literature is
incorporated in these figures.

\section{Conclusions \label{SECT_CONC}}

The results presented in this paper show that it is possible to
classify remarkably well the \hip periodic variable stars into types
that reflect their stellar physical properties. As detailed in
Sect.~\ref{SECT_ERR_ANA}, the main confusion cases are quite well
understood. They originate from any of: (1) problems in extracting the
correct period in the case of eclipsing binaries and \elli variables,
(2) real similarities between different types of Cepheid stars,
(3) a known, true difficulty for disentangling different types of blue
variables, in particular the \spb and \acv stars. In any case, as
seen in Fig.~\ref{FIG_CM} the classification errors related to these
confusion cases are generally below the 10\% level. This figure also
shows that they are only a handful of additional confusion cases.

Similarly good results are obtained with the random forest methodology
as with the multi-stage approach.  Important advantages of random
forest include a very useful attribute ranking method and a simple
set-up and tuning. Is it also surprisingly robust to the presence of
irrelevant or highly correlated attributes. The multi-stage approach
allows a controlled selection of a particular classification algorithm
and of a different attribute set at each node. Such choices can be
useful in specific cases, but they also require more extensive and
time-consuming optimization work.

Experience with various classification methods, random forest,
multi-stage and other alternative methods, suggest that significant
improvements are unlikely to come from better classification
algorithms. Important progress rather can be expected through the
introduction of new attributes which better reflect features of the
physical processes responsible for the variability. They may even be
specifically designed to disentangle some known cases of
confusion. This is possible for example when additional independent
data such as colour light curves, or radial-velocity time series are
available.

In addition to presenting the first systematic automated
classification of the Hipparcos periodic variable stars, this paper
describes the construction of a homogeneous training set of periodic
stars. In a companion paper (Rimoldini et al. in preparation), this
training set is completed with non-periodic variable stars. The
complete training set can then be adapted to other surveys as a
starting point for further classification studies. Some challenging
topics, such as variability detection, period search reliability and
possible confusion between periodic and non-periodic types are
deferred to subsequent investigations.

\section*{Acknowledgments}

We warmly thank Andy Liaw, the author of the random forest R
implementation, for suggesting the optimisation procedure used in
Sect.~\ref{SECT_ANDY}. TL acknowledges support by the Austrian Science
Fund FWF under project P20046-N16.  We thank an anonymous referee for
a careful review of this paper.

\bsp

\label{lastpage}

\end{document}